\documentclass[12pt,a4paper,final]{iopart}

\usepackage{graphicx}
\usepackage[breaklinks=true,colorlinks=true,linkcolor=blue,urlcolor=blue,citecolor=blue]{hyperref}

\expandafter\let\csname equation*\endcsname\relax
\expandafter\let\csname endequation*\endcsname\relax
\usepackage{amsmath}
\usepackage{amssymb}

\usepackage{multirow}
\usepackage{adjustbox}

\usepackage{subcaption,siunitx,booktabs}

\begin{document}

\title[On self-play computation of equilibrium in poker]
{On self-play computation of equilibrium in poker}

\author{Mikhail Goykhman}
\address{Racah Institute of Physics, Hebrew University of Jerusalem,\\
Jerusalem, 91904, Israel}
\ead{\mailto{michael.goykhman@mail.huji.ac.il} \footnote{\mailto{goykhman89@gmail.com}}}

\providecommand{\keywords}[1]{\textbf{\textbf{Keywords.\\}} #1}

\begin{abstract}

We compare  performance of the genetic algorithm and the counterfactual
regret minimization algorithm in computing the near-equilibrium strategies in the simplified poker games.
We focus on the von Neumann poker and the simplified version of the Texas Hold'Em poker,
and test outputs of the considered algorithms against analytical expressions defining
the Nash equilibrium strategies. We comment on the performance of the studied algorithms against
opponents deviating from equilibrium.

\end{abstract}


\section{Introduction}

The subject of game theory was given a rigorous mathematical foundation
by von~Neumann and Morgenstern \cite{Neumann1944}. From the earliest days of the
mathematical game theory the game of poker has been used as a testing ground for
the formal theory. In \cite{Neumann1944}
a simple heads up (two players) poker game, now usually referred to as the von~Neumann poker,
was proposed as an analytically tractable
variant of a game of imperfect information and a poker-like betting structure. This game
has been solved exactly, in the sense that equilibrium strategies for both of the players have been found
\cite{Neumann1944}. In the game equilibrium no player can expect an improvement in their
performance by unilaterally deviating from the equilibrium strategy.

Two players in the von Neumann poker are assigned asymmetric betting positions.
One of the players, sitting in the first position, can place the first bet, which the other
player, sitting in the second position, would consider to call. Such an asymmetry between the playing
positions resembles the betting sequence in the real poker games, usually defined
by the position of the players w.r.t. the player 'on the button' (dealer).
It has been shown in \cite{Neumann1944}
that under the rules of the von Neumann poker the player who
can place the first bet has an advantage, given the equilibrium play, while the equilibrium strategy
for the player in the second position minimizes their disadvantage (von Neumann
poker is a zero-sum game, that is, winnings of one player are equal in absolute value
to the losses of the other player).
In general, existence of such equilibrium strategies,
in two-person zero-sum games, for both players is guaranteed by the minimax theorem \cite{Neumann1928}.
Typically in the 
poker games the position of the players alternates between the rounds of play. Therefore on average
over many rounds of play both players have zero expected winnings when they play at equilibrium.

The equilibrium solution to the two-person zero-sum games also belongs to the class
of game-theoretic strategies known as the Nash equilibrium. However the Nash equilibrium
can be formulated more generally for $n$-person non-zero sum games,
as an $n$-tuple of strategies for $n$ players such that no player can improve their (average)
payoff by unilaterally deviating from the equilibrium strategy \cite{Nash1950}. Usually to solve a heads up
poker game (such as the heads up Texas Hold'Em) means to find the Nash equilibrium strategy.
The motivation behind identifying optimal strategy and the Nash equilibrium
(despite forfeiting profitable exploitation of a possibly sub-optimal non-equilibrium play of the opponent) is that 
the poker agent which plays according
to the Nash equilibrium strategy will have a guaranteed positive expectation value over many rounds of game
against the opponent who does not follow the Nash equilibrium, while remaining immune to
being exploited itself.

The earliest attempt to find an approximate Nash equilibrium in the full game of the limit Texas Hold'Em
was described in \cite{Billings2003}, see \cite{Billings2006,Rubin2011} for the review and discussion of the history of
development of computerized poker.
Recent advances have allowed to create artificial intelligence poker agents capable of playing
at the level exceeding top human professionals, both in the limit \cite{Bowling2015} and no-limit
\cite{Moravc2017,Brown2017} Texas Hold'Em.  Notice that when there are more than two players,
following the Nash equilibrium strategy might end up being sub-optimal, if more than one opponent deviates
from the Nash equilibrium strategy. Even assuming the goal of finding the Nash equilibrium,
the methods usually applied to achieve this goal for the heads up games might have limitations in the multi-player game,
see \cite{Gibson2014} for recent developments and review of the progress.

\vspace{0.5cm}

In this paper we are interested in comparing various approaches to 
calculation of the Nash equilibrium in poker games through the self-play simulations and training. The specific games which we focus
on are the von Neumann poker (as defined in subsection \ref{introducing_von_neumann}) and the flop poker
(as defined in section \ref{sec:flop_poker}). We suggest the flop poker as an immediate
upgrade of the von Neumann poker, which retains some simplistic features
of the latter, yet adds to it realistic poker aspects. Specifically, unlike the von Neumann poker, the flop poker is being played with the actual $52$-card deck,
and has the hands and the community cards layout resembling the Texas Hold'Em (this game is also analogous
to the pre-flop Texas Hold'Em game of \cite{Barone1999,Selby1999}).
The essential distinction between the von Neumann and the flop poker
is that in the former the strength of the players private hands is determined unambiguously, while
in the latter, just like in the real poker, in the case when showdown occurs, the strength of the final hands
depends on the dealt community cards, and
as a result any private hand can end up being the strongest.

The game of poker, where the players take turns to act, is straightforward to represent
in an extensive form, that is, in terms of the game tree.
One way to find an equilibrium for the extensive (two person) game is to first bring it
to a matrix form (in general known as a normal form), in which the payoff
is a bi-linear function of two players strategies.
The matrix game can be further recast as a linear program, and solved, for instance,
using the simplex algorithm. However for most of the games with even small number of nodes
in their game tree
this method is computationally unfeasible.
Indeed, strategies of the players in the most general form
live in the space which is a direct product of the spaces of pure strategies at each decision node of the game tree.
Therefore the size of the strategy vector is exponential in the size of the game tree. For instance,
a discrete version of the von Neumann poker, in which each player receives a random integer number
in the range $1\dots 100$, has the strategy space for each player of the dimension $2^{100}\simeq 10^{30}$. 
The sequence form of the game has been proposed to circumvent this 
issue, allowing the representation linear in size of the game tree, which can then be solved
in the framework of linear programming \cite{Koller1994}. The equilibrium pre-flop poker strategies
can then be found \cite{Selby1999,Billings2003}.

Alternative methods to find equilibrium game strategies have been used for games with the large game tree.
The state-of-the-art method used at the core of the solution of the real poker games
(limit and no-limit Texas Hold'Em \cite{Bowling2015,Moravc2017,Brown2017}) is the
counterfactual regret minimization algorithm (CFR)  \cite{Zinkevich2007},
see \cite{Neller2013} for the review of the CFR and its predecessor,
the regret matching algorithm \cite{Hart2000}.
\footnote{Other issues have to be addressed, such as reducing the size of the game via an efficient abstraction.
One needs to resolve the issue of the strategy becoming exploitable due to the nature of the game
abstraction \cite{Shi2001,Billings2003}. We will not be discussing these issues in this paper.}
The CFR algorithm is capable of finding near-equilibrium strategy of a heads up
two-person zero-sum game with incomplete information, known as the $\epsilon$-Nash
equilibrium.

Another approach which has been applied to solve
poker games is based on the evolutionary optimization algorithms, such as genetic
algorithms \cite{Holland1975}, and evolutionary programming \cite{Fogel1966}.
In general application of evolutionary optimization
(in particular to the game of poker) should be taken with caution. Several caveats have been pointed
out in the literature, such as the bias created by the sub-optimal strategies,
which happened to be lucky in the given round of evolution, which results in warping of the algorithm output
\cite{Fogel2000}. The issue is that under certain criteria of evolutionary algorithm,
such as selection of a very small number of the most fit parents (high selection pressure), one
might end up picking the lucky strategies (found at the far end
of the performance distribution) rather than the strategies with the highest expected
value. Another subtlety of applying evolutionary optimization to selection of the optimal
poker strategies is related to the well-known non-transitive nature of poker (see \cite{Chen2006}
for a recent discussion). Non-transitive games (such as the simple game of Rock-Paper-Scissors)
can end up evolving cyclically instead of converging
to equilibrium \cite{MaynardSmith1982}.

One of the earlier approaches which applied evolutionary optimization methods to
devise a poker playing agent is given in \cite{Barone1999}. In that paper the simplified
poker game was considered, in which two players in the setup of the limit Texas Hold'Em make the sequence of bets after receiving private hands of two cards, after which (if no player folds) five community cards are dealt, and the players make the best five-card hand
out of two private cards and five community cards. The two-card hands received by the players
have been assigned the ranks of strength, related to the probability to win with those hands.
The game decisions to bet, call, raise, or fold have been determined probabilistically as heuristic
functions of the hand rank. These functions were defined by a small
set of parameters, which subsequently were optimized evolutionary. 
One of the main points of \cite{Barone1999} was to demonstrate how evolutionary selection
can optimize the game against the given opponent.
A more recent example
\cite{Quek2009} uses the loss minimization genetic algorithm, and the hand
strength card abstraction, to simultaneously co-evolve
two players playing heads up Texas Hold'em poker game. See also \cite{Carter2007}
for some further applications of the evolutionary optimization to poker games,
including the Kuhn poker \cite{Kuhn1950}.

The goal of this paper is to compare performance of the genetic algorithm
and the counterfactual regret minimization algorithm in the self play of two poker agents,
both of which start with random strategies without any pre-programmed knowledge of the optimal play.
For our purposes we consider the games of the von Neumann poker and the flop poker.
We demonstrate that for the task of calculating the near-equilibrium
strategy in the flop poker, the CFR algorithm in general has less noisy output
and better
convergence properties than the genetic
algorithm, consistently with the popularity of CFR in developing the top
poker playing agents. At the same time both the genetic algorithm and the CFR algorithm
perform similarly well in determining the near-equilibrium
strategies in the von Neumann poker. We also point out that the simple CFR algorithm
randomly finds one out of many equilibrium strategies for the second player in the
von Neumann poker, while the genetic algorithm typically finds the second player's equilibrium strategy which
is the most exploitative of the first player's deviations from the equilibrium.

The rest of the paper is organized as follows. In section \ref{von_neumann} we review
the von Neumann poker. We start by discussing how the von Neumann poker can be motivated
as a simplification of the real card poker games. We then proceed to deriving the equilibrium solution 
to the von Neumann poker. In section \ref{sec:von_neumann_evolve} we discuss
the results of application of the genetic algorithm to the problem
of calculating the near-equilibrium strategies in the von Neumann poker. We review
general principles of the genetic algorithm which we will also apply to calculate
equilibrium in the flop poker.  In section \ref{CFR_von_neumann} we start by reviewing
the CFR algorithm, focusing on its application to the
von Neumann poker. We then discuss the results of applying the CFR algorithm to compute the
$\epsilon$-Nash equilibrium strategies in the von Neumann poker.
In section \ref{sec:flop_poker} we introduce the flop poker, and derive expressions
which determine its Nash equilibrium strategies. We calculate the near-equilibrium
strategies in flop poker using the genetic algorithm in section \ref{sec:flop_poker_evolution},
and using the CFR algorithm in section \ref{sec:flop_poker_cfr}.
We discuss our results in section \ref{sec:discussion}. In appendix we provide
details of our poker hand evaluator.

\section{Von Neumann poker}
\label{von_neumann}

In this section we are going to discuss the two-person zero-sum game known as von~Neumann poker.
This game was originally formulated in \cite{Neumann1944} (a similar game, sometimes
referred to as Borel poker, was introduced in \cite{Borel1938}),
as an example of a game which retains some of the essential
features of the poker games, yet is simple enough to be solved exactly.
We review how one can arrive at the von Neumann poker starting
from the real poker games in subsection~\ref{introducing_von_neumann}.

One of the most straightforward ways to find equilibrium solution to the von~Neumann
poker uses the principle of indifference.
This method determines the optimal equilibrium strategy which is also admissible,
that is, maximally (among all of the possible equilibrium strategies) exploitative
of the opponent's deviations from their equilibrium,
see \cite{Ferguson2003} for a recent review and
developments. It is known
that while the first player's strategy in the von Neumann poker is unique, the second player has a continuum of
optimal strategies, all of which are equilibrium strategies, resulting in the same value of the game \cite{Neumann1944}.
In subsection \ref{von_neumann_solution} we provide a comprehensive
derivation which finds all the equilibrium solutions to the von Neumann poker.
The content of subsection \ref{von_neumann_solution} does not provide any new results,
but rather presents our perspective on the von Neumann poker.

We are planning to take advantage of knowing the exact equilibrium solutions to the von Neumann poker
to test outcome of the genetic algorithm and the CFR algorithm in
 computing the near Nash equilibrium strategies, as
discussed below in sections \ref{sec:von_neumann_evolve}
and \ref{CFR_von_neumann} of this paper. In particular we point out that the evolutionary optimized strategy
typically approaches the admissible equilibrium strategy, as defined above.
On the other hand the 
counterfactual regret minimization finds randomly one of the equilibrium strategies, which does not take
any advantages of the possible opponent's deviations from their equilibrium.

\subsection{Introducing von Neumann poker}
\label{introducing_von_neumann}

The idea behind the von Neumann poker originates from the desire to retain essential characteristic features of
the variety of poker games, while lighting up the
complicated specific rules of the actual card games \cite{Neumann1944}.
The resulting two-player game is a game of incomplete information, which involves rounds of betting,
during which players can check, bet, call, or fold (raises are not allowed in the simplest
version of the game), and therefore resembles the kind of games which are usually defined as poker.

We begin by reviewing how one can arrive at the von Neumann poker by starting with
the real variants of poker \cite{Neumann1944}.
Each player in the typical game of poker receives their own private cards, which can be used to compose
a hand (we discuss poker which is played with 52-card deck).
Then several rounds of betting occur, which can result in all but one player
folding their hands. The remaining player then collects the entire pot, while cards are not revealed,
and no hands are compared in strength. However, if players check, or if at least one player
calls,
then a showdown occurs, in which case the best hand among the remaining players wins the pot.

Apart from additional criteria, such as the `tells' of the other players, their betting patterns, round of game, and the position of the given player in the
game, etc., the decision of how to act in each situation is defined by the hand which the player holds
(in general, by the cards which the player can claim).
Each poker strategy is defined by the set of prescriptions of how to act with each particular
poker hand. These prescriptions usually amount to assigning the probabilities to various
actions which the players will follow with each possible hand and in each possible situation.
We will provide the derivation of optimal strategies in the case of von Neumann poker below in this section.
For now we focus on discussing what possible hands the player can make.

All possible hands can be ranked from the weakest one to the strongest one. Consider, for instance,
the common poker variant in which a hand is a set of five cards. These cards can all
be privately held by a player (all five cards are private cards, as in the Five-card draw poker), or can be composed by the player's private cards and the community cards (as in the Texas Hold'Em game).
Regardless of these distinctions in the specific poker rules, we can always rank the hands of five cards from
the weakest one (which is beaten by all the other five-card hands), to the strongest one (which beats
all the other five-card hands).

The most broad classification of strength of the five-card poker hands divides all hands
into nine categories, from the highest to the lowest,
Straight Flush (including Royal Flush), Four of a Kind, Full House, Flush, Straight, Three of a Kind, Two Pair, Pair,
and High Card. With this ranking of poker hands it is not infrequent that
two players will end up having the same rank, for instance, two players might end up each having a Pair.
We know that the highest Pair wins (cards are ranked, in increasing strength, from Deuces to Aces),
or in case when both players have the same Pair, then the kickers (the remaining
three cards in the hand) are compared in strength, and the hand with the highest kicker wins.
That is, poker hands can be ranked more finely in strength than the nine groups listed above.

We now review the precise way to rank all possible five-card poker hands. There are
\begin{equation}
\label{52_choose_5}
{\cal N}={52\choose 5}=2,598,960
\end{equation}
ways to deal five cards out of 52-card deck. However many of these hands are equal in strength,
for instance, Four of a Kind, composed of four Aces and a King has the same strength regardless
of one of the four possible suits which the King might have.
Once the equivalent hands are grouped together (in other words, the suit degeneracy is factored out), the actual number of distinct hands becomes
equal to 7462, as can be seen in Table \ref{poker_ranks}, where all possible hands,
the number of distinct ranks, their degeneracies,
and the total number of hands are listed.

\begin{table}[h!]
\caption{Ranking of five-card poker hands.\label{poker_ranks}}
\begin{center}
\begin{tabular}{ l l l l}
\hline
 {\rm Name} & {\rm Distinct hands} & {\rm Degeneracy} & {\rm Total}\\ [0.5ex] 
 \hline
 {\rm Straight Flush} & 10 & ${4\choose 1}=4$ & 40 \\ 
 {\rm Four of a Kind} & ${13\choose 1}{12\choose 1}=156$ & ${4\choose 1}=4$ & 624 \\ 
 {\rm Full House} & ${13\choose 1}{12\choose 1}=156$ & ${4\choose 3}{4\choose 2}=24$ & 3,744 \\ 
 {\rm Flush} & ${13 \choose 5}-10=1,277$ & ${4\choose 1}=4$ & 5,108 \\ 
 {\rm Straight } & 10 & $4^5-4=1,020$ & 10,200 \\ 
 {\rm Three of a Kind} & ${13\choose 1}{12\choose 2}=858$ & ${4\choose 3}{4\choose 1}^2=64$ & 54,912 \\ 
 {\rm Two Pair} & ${13\choose 2}{11\choose 1}=858$ & ${4\choose 2}^2{4\choose 1}=144$ & 123,552 \\ 
 {\rm Pair} & ${13\choose 1}{12\choose 3}=2,860$ & ${4\choose 2}{4\choose 1}^3=384$ & 1,098,240 \\ 
 {\rm High Card } & ${13\choose 5}-10=1,277$ & ${4\choose 1}^5-{4\choose 1}=1,020$ & 1,302,540 \\ 
 \hline
 {\rm Total} & 7,462 & --- & ${52\choose 5}=2,598,960$ \\
 \hline
\end{tabular}
\end{center}
\end{table}

Let us use index $i$, taking values from $1$ to $7462$, to label distinct five-card hands,
where $i=1$ is the Ace-high Straight Flush (also known as the Royal Flush), and $i=7462$ is 
the High Card $7\,5\,4\,3\,2$.
Denote degeneracy of
each hand as $d(i)$, for instance degeneracy of each Straight Flush is $d(i_{{\rm s.f.}})=4$, $i_{{\rm s.f.}}=1,\dots,10$.
Then the probability to get hand $i$ is given by 
\begin{equation}
\label{h_of_i_def}
h(i)=\frac{d(i)}{{\cal N}}\,,
\end{equation}
where total number of dealings ${\cal N}$ was defined in (\ref{52_choose_5}).
Since degeneracies vary, 
probability of getting hand $i=1,\dots,7462$ varies, depending on which of the nine
groups in Table~\ref{poker_ranks} the $i$ is in.

The proposal made by von Neumann was to remove this complication of having
non-uniform probabilities of various hands.
Instead, von Neumann suggested to consider the game in which each player is privately dealt one of
$S$ numbers, with the uniform probability distribution, $h(i)=1/S$, assigned to each number $i=1,\dots,S$.
The strategy of the player is then to be determined by the number $i$.

Von Neumann subsequently takes the
continuous limit, and considers the game where each player is dealt a number from $[0,1]$, with the uniform
probability distribution.
The specific rules of the game are follow. There are two players, which we call Player and Dealer.
\footnote{This naming convention distinguishes the order of betting, the Player gets to act first.
Both Player and Dealer will be referred to as players, with the lower-case {\it p}.}
Before the round of game starts, each player puts
ante $a$ into the pot. Then each player is dealt a uniformly-drawn random number from $[0,1]$.
The following round of betting subsequently takes place.
Player can either check, or bet $B$ (clearly only ratio of $B/a$ matters, and all the results
are expected to be invariant w.r.t. simultaneous rescaling of $a$ and $B$).
If the Player checks, then the showdown occurs, in which
case the player with higher number wins the pot, $P=2a$.
\footnote{When discussing the discrete version of the hands labeling we chose the smallest
index $i=1$ to denote the strongest hand (Royal Flush). In our discussion of von Neumann
poker the higher number will stand for the stronger hand. We hope this will not cause a confusion,
since the section on the solution to the von Neumann poker can be read separately.}
If the Player bets, then the decision is
passed on to the Dealer. If the Dealer folds, then the pot, $P=2a$, is won by the Player. If the Dealer calls,
then the showdown occurs, and the player with the highest  number wins the pot, $P=2a+2B$.
The problem is to derive optimal strategies for both players, that is, to find with what hands the Player
should bet, and with what hands the Dealer should call when facing the bet (more generally, what is the probability
with which Player/Dealer should bet/call with each possible hand). This is the game which we will be studying in this section.

\subsection{Equilibrium in von Neumann poker}
\label{von_neumann_solution}

To solve the game usually means to find the Nash equilibrium. Nash equilibrium
is defined as a strategy which is the best game against itself: no player will be better off
by unilaterally deviating from the Nash equilibrium, if everyone else is playing according to the Nash equilibrium
strategy. In other words, Nash equilibrium also possesses the feature of being a non-exploitable strategy;
if Player 1 deviates from Nash equilibrium then Player 2 will be exploiting Player 1, and increase
their (average) payoff.
Generally identifying Nash equilibrium and optimal game strategy is not always correct. If one of the players
is known not to play by the Nash strategy, then the correct strategy is to maximally exploit that player, increasing one's own
payoff as a result. We will assume that it is a common knowledge \cite{Lewis1969} that the players are rational, and therefore
everyone will play according to the Nash equilibrium strategy. Then for each individual player it is optimal to also follow
the Nash equilibrium strategy.\footnote{As discussed in Introduction, solving the actual games of poker numerically
 is usually done by finding the strategy which is as close as possible to the Nash equilibrium.
For large games such as poker this is also technically easier
than trying to develop an algorithm which attempts to observe and exploit weaknesses of its opponent.
The idea behind a `conservative' Nash play is that the (human) opponent will not be able to figure out the Nash equilibrium strategy nearly as well as the AI poker agent, and therefore
will end up being worse off in the long run anyway. This is among the reasons why the effort in constructing
poker agents has been focused on finding the Nash equilibrium.}

Motivated by this assumption we proceed to derive the Nash equilibrium for the defined game of the von Neumann poker.
Denote the number dealt to the Player as $x$, and the number dealt to the Dealer as $y$. Player follows the strategy
which prescribes the probability to bet, $p(x)$,
and the probability to check, $1-p(x)$, with the hand $x$. Dealer follows the strategy which prescribes
the probability to call (if facing the bet), $q(y)$, or to fold, $1-q(y)$, when holding the hand $y$. The expected
gross winnings of the Player holding hand $x$, and the Dealer holding hand $y$, are
\begin{align}
E_1(x)&=p(x)\,e_1(x)+Px\,,\label{Player_expected_winning}\\
E_2(y)&=q(y)\,e_2(y)+P\left(y-\int _0^ydx\,p(x)\right)\,.\label{Dealer_expected_winning}
\end{align}
Here we have introduced the functions
\begin{align}
e_1(x)&=P(1-x)+B\,\int _0^x dy\,q(y)-(P+B)\,\int _x^1 dy \,q(y)\,,\label{e1_def}\\
e_2(y)&=(P+B)\,\int _0^y dx\,p(x)-B\,\int _y^1dx\,p(x)\,,\label{e2_def}
\end{align}
which serve an important purpose, that will explained momentarily.
We denoted $P=2a$, which is the pot comprised by the initial antes put in by the Player
and Dealer, before they receive their hands $x$ and $y$.

We are working in the pot framework,
where $P$ is considered a `sunk cost', which means that in the calculation of the expected gross winnings  
(\ref{Player_expected_winning}), (\ref{Dealer_expected_winning}), forfeiting the $P=2a$
by folding the hand, or not
winning it during a showdown, was not incorporated as a loss. 
This way the game is $P$-sum, rather than zero-sum,
\begin{equation}
E_1+E_2=P\,,
\end{equation}
where the total gross winnings of the Player and the Dealer are
\begin{equation}
\label{gross_winnings}
E_1=\int _0^1 dx\, E_1(x)\,,\qquad E_2=\int _0^1dy\, E_2(y)\,.
\end{equation}
The net winnings are
\begin{equation}
\label{net_winnings}
E_{1,2}^{{\rm net}}=E_{1,2}-\frac{P}{2}\,.
\end{equation}
Notice that working in the pot framework is completely equivalent to working in the zero-sum framework,
in the latter case the expected winnings would also be the net winnings, but the solution for the equilibrium
strategies would, of course, be exactly the same as the one derived working in the pot framework.
In other words, using the pot framework is an optional choice, and is used for the purpose of convenience only.
(Usually in poker the pot framework is also convenient to calculate the pot odds, the value which is to be compared
with the probability to win the game, in order to determine whether the bet is worth a call.)

The functions $e_{1}(x)$, $e_2(y)$, introduced in (\ref{e1_def}), (\ref{e2_def}), define the optimal
betting strategies for the Player and the Dealer. Consider, for instance, the Player.
If $e_1(x)<0$, then according to (\ref{Player_expected_winning}) the Player will maximize
their expected winnings by choosing $p(x)=0$, that is, always checking when holding $x$. Similarly, if $e_1(x)>0$,
the Player is the best off by playing with $p(x)=1$, that is, always betting when holding $x$. If $e_1(x)=0$,
then the Player is indifferent to choosing a specific $p(x)$. Notice that $e_1(x)$,
which influences the optimal Player's play,
is determined by the Dealer's strategy $q(y)$. Therefore the Dealer's strategy can be exploited by the Player.
The Nash equilibrium is achieved when the strategies are not exploitable, that is, when deviating from that strategy
unilaterally leaves the player worse off.

To find the Nash equilibrium in the von Neumann poker we begin by noticing that 
\begin{equation}
\frac{de_2}{dy}=(P+2B)\,p(y)\geq 0\,.\label{e2_derivative}
\end{equation}
Combined with the observations (we exclude trivial game in which Player never bets,
that is, $p(x)\equiv 0$ for all $x\in [0,1]$)
\begin{equation}
e_2(0)=-B\,\int _0^1dx\,p(x)<0\,,\quad e_2(1)=(P+B)\,\int _0^1 dx\,p(x)>0\,,
\end{equation}
expression
 (\ref{e2_derivative}) implies that $e_2(y)$ is a non-decreasing function, which goes from a negative value at $y=0$,
to positive value at $y=1$. Let us denote $[x_1,x_2]$ to be the interval where $e_2(y)$
passes through zero (since $e_2(y)$ in general goes through zero via an interval rather than
a single point, although it might be that $x_1=x_2$). Consequently the optimal strategy for the Dealer
when $y$ is outside of the interval $[x_1,x_2]$, is given by
\begin{equation}
\label{y_premliminary_solution}
q(y) = \left \{
  \begin{aligned}
    &0  && \ y\in [0,x_1) \\
    &1 && \ y \in (x_2,1]\,.
  \end{aligned} \right.
\end{equation} 
Since $e_2(y)=0$ for $y\in [x_1,x_2]$, the Dealer is indifferent to their strategy $q(y)$ for $y\in [x_1,x_2]$.
Let us denote
\begin{equation}
c=\int _{x_1}^{x_2}dy\,q(y)\,.\label{c_def}
\end{equation}
To complete the Nash equilibrium solution for the Dealer we need to find the optimal value of $c$.

We now switch our focus to the Player's strategy. According to the definition of $x_{1,2}$
and due to (\ref{e2_derivative}), we know that
$p(x)=0$ for $x\in (x_1,x_2)$. From (\ref{e2_def}) due to $e_2(y)=0$, $y\in (x_1,x_2)$, we also obtain
\begin{equation}
(P+B)\, \int _0^{x_1}dx\, p(x)=B\,\int _{x_2}^1dx\,p(x)\,.
\label{px_balance_condition}
\end{equation}

Using the solution (\ref{y_premliminary_solution}), (\ref{c_def}) for the Dealer we can calculate the Player's
function $e_1(x)$, defined in (\ref{e1_def}). We focus now on the regions outside of the interval $[x_1,x_2]$,
\begin{equation}
\label{e1_premliminary_solution}
e_1(x) = \left \{
  \begin{aligned}
    &-Px+P-(P+B)(1+c-x_2)  && \ x\in [0,x_1) \\
    &2B x+B(c-1-x_2) && \ x \in (x_2,1]\,.
  \end{aligned} \right.
\end{equation} 
We are searching for the solution in which the Player bets at least with some hands.
Therefore we expect that the Player will bet at least in some region near $x=1$,
where it has the strongest hands.
From (\ref{e1_premliminary_solution}) we see that $e_1(x)$ is monotonically increasing
when $x\in (x_2,1]$. Since we know that $p(x)=0$ for $x\in (x_1,x_2)$, and $p(x)$ in equilibrium is
determined by the  sign of $e_1(x)$, then $e_1(x)<0$ for $x\in (x_1,x_2)$, and therefore $e_1(x_2)=0$.
Therefore from (\ref{e1_premliminary_solution}) we obtain
\begin{equation}
x_2=1-c\,.\label{x2_dependence_on_c}
\end{equation}

From (\ref{e1_premliminary_solution}) we also observe that $e_1(x)$ is monotonically
decreasing in $[0,x_1)$. It is unclear what is the sign of $e_1(0)$, that is,
whether the Player will bet in the vicinity of $x=0$. It can be shown that the solution where
the Player never bets for small $x$ is trivial, that is, the Player would never bet and the Dealer would never
have to call.
\footnote{Since we know that $e_1(x)<0$ for $x\in (x_1,x_2)$,
then the assumption $e_1(0)<0$ actually implies that $e_1(x)<0$ for $x$
in the entire region $[0,x_2)$,
because $e_1(x)$ is monotonically decreasing in $[0,x_1)$.
Then $p(x)=1$ for $x\in (x_2,1)$ and $p(x)=0$ otherwise. Using (\ref{e2_def})
we observe that then $e_2(x_2+0)<0$, contrary to the definition of $x_2$. The only
way out in this case is to set $x_2=1$, which means that Player never bets.
} 
Therefore we proceed by assuming that $e_1(x)>0$. In that case, since $e_1(x)$
is monotonically decreasing in $[0,x_1)$ and (since $p(x)=0$ for $x\in (x_1,x_2)$)
in equilibrium we expect that $e_1(x)<0$ for $x\in (x_1,x_2)$, then $e_1(x_1)=0$.
Using (\ref{e1_premliminary_solution}) we then obtain
\begin{equation}
x_1=1-2\frac{P+B}{P}\,c\,.\label{x_2_dependence_on_c}
\end{equation}
The equilibrium solution for the Player is then 
\begin{equation}
\label{player_solution}
p(x) = \left \{
  \begin{aligned}
    &1  && \ x\in [0,x_1)\ {\rm and} \ x\in (x_2,1]\\
    &0 && \ x \in (x_1,x_2)\,.
  \end{aligned} \right.
\end{equation} 
Using (\ref{x2_dependence_on_c}), (\ref{x_2_dependence_on_c}),
and (\ref{player_solution}) in
(\ref{px_balance_condition}) we obtain
\begin{equation}
\label{c_solution}
c=\frac{P(P+B)}{PB+2(P+B)^2}\,.
\end{equation}
Plugging (\ref{c_solution}) back into (\ref{x2_dependence_on_c}), (\ref{x_2_dependence_on_c}) 
we obtain
\begin{equation}
\label{x1_x2_solution}
x_1=\frac{PB}{PB+2(P+B)^2}\,,\qquad  x_2=\frac{2(P+B)^2-P^2}{PB+2(P+B)^2}\,.
\end{equation}

To complete the solution we need to specify the Dealer's strategy for $y\in (x_1,x_2)$.
We already know that the Dealer's strategy is given by (\ref{y_premliminary_solution})
outside of the interval $[x_1,x_2]$, and that within that interval the strategy 
is constrained by (\ref{c_def}), where $c$ is given by (\ref{c_solution}).
The last requirement which we need to impose to make the solution consistent is
to study how $q(y)$ affects $e_1(x)$ when $x\in (x_1,x_2)$, the latter has been excluded so far from (\ref{e1_premliminary_solution}).
To ensure that the solution is consistent, that is, that $p(x)=0$ for $x\in (x_1,x_2)$,
we need to have $q(y)$ such that $e_1(x)<0$ for $x\in (x_1,x_2)$. One such solution is
\begin{equation}
\label{y_x1x2_solution}
q(y) = \left \{
  \begin{aligned}
    &0  && \ y\in (x_1,y_0) \\
    &1 && \ y \in (y_0,x_2)\,,
  \end{aligned} \right.
\end{equation}
where
\begin{equation}
\label{y0_defintion}
y_0=x_2-c=\frac{B(3P+2B)}{PB+2(P+B)^2}\,.
\end{equation}
However, as can be shown, this is not the only solution which 
ensures that  $e_1(x)<0$ for $x\in (x_1,x_2)$, and satisfies the 
constraints (\ref{c_def}), (\ref{c_solution}). The solution (\ref{y_x1x2_solution}), (\ref{y0_defintion})
is usually referred to as admissible equilibrium solution. Of all the equilibrium
solutions for the Dealer the solution (\ref{y_x1x2_solution}), (\ref{y0_defintion})
takes the most advantage if the Player deviates from their own equilibrium strategy
(\ref{player_solution}), (\ref{x1_x2_solution}). 

Finally, using (\ref{net_winnings}) we find that the net winnings per round for the Player and the Dealer are
\begin{equation}
E_1^{{\rm net}}=\frac{P}{2}\frac{PB}{PB+2(P+B)^2}\,,\quad E_2^{{\rm net}}=-\frac{P}{2}\frac{PB}{PB+2(P+B)^2}\,.
\end{equation}
The game favors the Player, which justifies the assumption made earlier, that the Player will
prefer to bet (and have a positive average payoff $E_1^{{\rm net}}>0$) than play a trivial game
(and have a zero average payoff $E_1^{{\rm net}}=0$).

\section{Evolutionary selection of strategies in the von Neumann poker}
\label{sec:von_neumann_evolve}

In this section we are going to apply the genetic algorithm
to calculate near-equilibrium strategies in the von Neumann poker game. Specifically we set up numerical 
simulations with
the goal to find out whether
evolutionary optimization will converge to the known Nash equilibrium solution
reviewed in section \ref{von_neumann}.
We will see that evolutionary optimization typically finds the Dealer's
solution (\ref{y_x1x2_solution}), (\ref{y0_defintion})
which is both equilibrium and admissible. This is because during the evolutionary optimization
the Dealer's equilibrium strategy which is admissible will perform better, since it will
take the most advantage of the mutated Player's strategies, which deviate from the Player's equilibrium.

In order to make
the problem numerically tractable we will consider the discrete version of the von Neumann
poker, in which each player is dealt a number $i,j=1,\dots, 100$, where the highest number
wins in case of showdown. The analytical solution described in section \ref{von_neumann}
will be used as an approximation to the discrete case studied in this section. 

We begin by reviewing the principles of evolutionary optimization and genetic algorithms in
subsection \ref{review_evolutionary_programming}.
The genetic algorithm described in subsection \ref{review_evolutionary_programming},
although illustrated on the example of the discrete von Neumann poker, 
will also be adapted in section \ref{sec:flop_poker_evolution} to calculate the near-equilibrium strategy in the flop poker.
We describe our results in subsection
\ref{results_evolutionary_von_neumann}.

\subsection{Review of the genetic algorithm}
\label{review_evolutionary_programming}

Our goal is to find optimal strategies for the Player and the Dealer. The Player's strategy 
is the vector $V_P$ of length $M$ (where $M=100$ in the discrete version of the von Neumann
poker which we are considering in this section), such that its entry $V_P(i)$ gives the probability with
which the Player bets if they hold the hand $i$. Similarly, the Dealer's strategy is the vector $V_D$
of length $M=100$, such that its entry $V_D(j)$ gives the probability with which the Dealer calls (if facing a bet)
if they hold the hand $j$.

In the framework of the genetic algorithm the strategy vectors $V_P$, $V_D$ are interpreted
as chromosomes, and their individual entries are assigned the role of genes.
Phenotypic manifestation of a gene is defined by the way the player acts in the game due to the value
of that gene. We start by initializing a population of $N$ Players and $N$ Dealers, with their
chromosomes prepared randomly. This population then evolves over $T$ rounds of evolution.

Evolutionary selection acts separately on the population of Players and the population
of Dealers. However, the selection of Players and Dealers is simultaneous, because the Players co-evolve with the Dealers.
We are going to assume that each entry of the strategy vectors (chromosomes) $V_{P,D}$, can be either 0 or 1,
that is, each decision is a pure strategy of either always acting, or never acting, where acting
stands for betting/calling for Player/Dealer. This is motivated partly by our prior insight into
the optimal strategy in the von Neumann poker, in other words, we know that the domain of solutions 
is a direct product
of pure binary strategies for each possible hand.

On the other hand, suppose that the equilibrium probability $V_P(i)$ for the Player to bet with some hand $i$
is not equal to 0 or 1. How is the evolutionary selection going to determine such a mixed strategy,
if every individual gene $V_P(i)$ is allowed to take the value of only either 0 or 1? We suggest to allow for the
possibility of a mixed strategy by taking the average of the chromosomes over the population. Such 
relation between the mixed strategy and the population polymorphism is known in evolutionary game theory,
and is based on the observation that playing against the opponent who has the mixed strategy $p$
is like being in a population of players where the fraction $p$ of the opponents have the pure strategy $V_P(i)=1$,
and the fraction $1-p$ have the pure strategy $V_P(i)=0$ \cite{MaynardSmith1982}.

We also assume that the Nash
equilibrium strategy which we are searching for is evolutionary stable strategy (ESS),
that is,
the population of ESS players cannot be invaded by the players who follow a different strategy
The concept of ESS is stronger than the Nash equilibrium \cite{MaynardSmith1982,MaynardSmith1973},
and will play a more important role in the discussion of the evolutionary optimization
of the flop poker strategies in section \ref{sec:flop_poker_evolution}.
With respect to the evolutionary stability the mixed strategy and the population polymorphism
are not interchangeable when more than two pure strategies are involved \cite{MaynardSmith1982}.

At each round $t=1,\dots,T$ of evolution $R$ games of poker are played.
In the beginning of each round of evolution the bankroll of all the players is
reset to the starting value $B_0$.
During each game Players
and Dealers are paired up randomly, and play one time. The wins and losses of each player are accumulated
through the $R$ rounds of game. After the games have been played, the Players and the Dealers are ranked by their
final bankroll $B_R$, and the fraction $\alpha$ of the best performers (as judged
by the highest bankroll $B_R$) are selected.
The $(1-\alpha) N$ of Players and the $(1-\alpha )N$
of Dealers are discarded. The $\alpha N$ Players/Dealers then produce $(1-\alpha) N$ offspring Players/Dealers.

Another way to select performers is to choose the ones who lost the least.
In that case the fit score of the member of the population is not changed when the money
is won, but decreases when the money is lost. In section \ref{sec:flop_poker_evolution} we consider such a selection
criterion, called loss minimization, in which the gains do not affect the fit score, while the losses
decrease it \cite{Holland1975}. This way the best score the Player/Dealer can get is zero. Therefore one can
unify Player and Dealer into one Participant object (because the game is zero sum, and the optimal
strategy for Participant object means playing optimally as a Player and optimally as a Dealer, with
the net result being zero), which will be assigned the role
of Player/Dealer randomly at each round of play.

The offsprings are produced in the following way.
Two Players/Dealers are selected randomly from the $\alpha N$ of the most fit players (parents), with the probabilities
proportional to their fit scores
(in the loss minimization framework of section \ref{sec:flop_poker_evolution}
the parents will be selected uniformly for the breeding, regardless of their fit score,
which is equal to zero at most). Two parents then produce one offspring.
Each gene $p_i$ in the offspring's chromosome slot $i$ is determined by the genes $p^{(1)}_i$
and $p^{(2)}_i$ of its parents. If $p^{(1)}_i=p^{(2)}_i$, then $p_i$ is set to $p_i^{(1)}=p^{(2)}_i$
with the probability $1-\pi$. However with the probability $\pi$ the gene will mutate to the flipped
value $(1+p_i^{(1)})\%2$.
On the other hand, if the parents have different genes at the slot $i$, $p^{(1)}_i\neq p^{(2)}_i$,
then the offspring will receive the gene $p_i^{(1,2)}$ from one of the parents randomly, with the probabilities
proportional to the fit scores of the parents (with equal probability in the loss-minimization framework). 

At the end of evolution the optimal strategy is calculated as follows. We sort all the Players and all the Dealers
according to their fit scores (final bankrolls). Then $\alpha N$ of the most fit Players and $\alpha N$
of the most fit Dealers are used to calculate the population average of the Player chromosome and the
Dealer chromosome. This way we allow for the possibility of obtaining a mixed strategy,
as discussed above.

\subsection{Results of evolutionary selection of the von Neumann poker strategies}
\label{results_evolutionary_von_neumann}

\begin{figure}
\begin{center}
 \includegraphics[width=12cm,height=12cm,keepaspectratio]{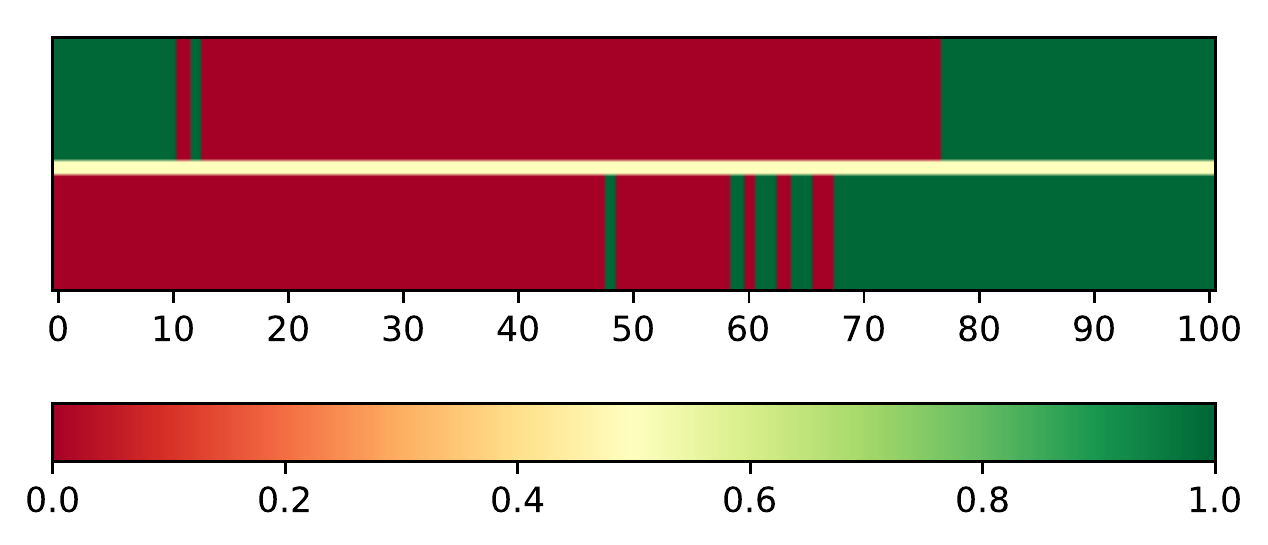}
 \caption{ \label{VN_evolve_strategy_B2_P2} 
 Von Neumann Player ({\bf top}) and Dealer ({\bf bottom})
 chromosome in subsection \ref{results_evolutionary_von_neumann},
 calculated as the final
  average of $\alpha=0.1$ of the most fit Players/Dealers in the population of $5000$ Players/Dealers, after $1000$ rounds of evolution,
 with $10000$ randomly paired games per Player/Dealer on each round. Players ante $a=1$ and bet $b=2$.
 Analytical prediction for the continuous game is $x_1=11$, $x_2=78$, $c=22$,  $y_0=56$.
 }
 \end{center}
\end{figure}

\begin{figure}
\begin{center}
 \includegraphics[width=12cm,height=12cm,keepaspectratio]{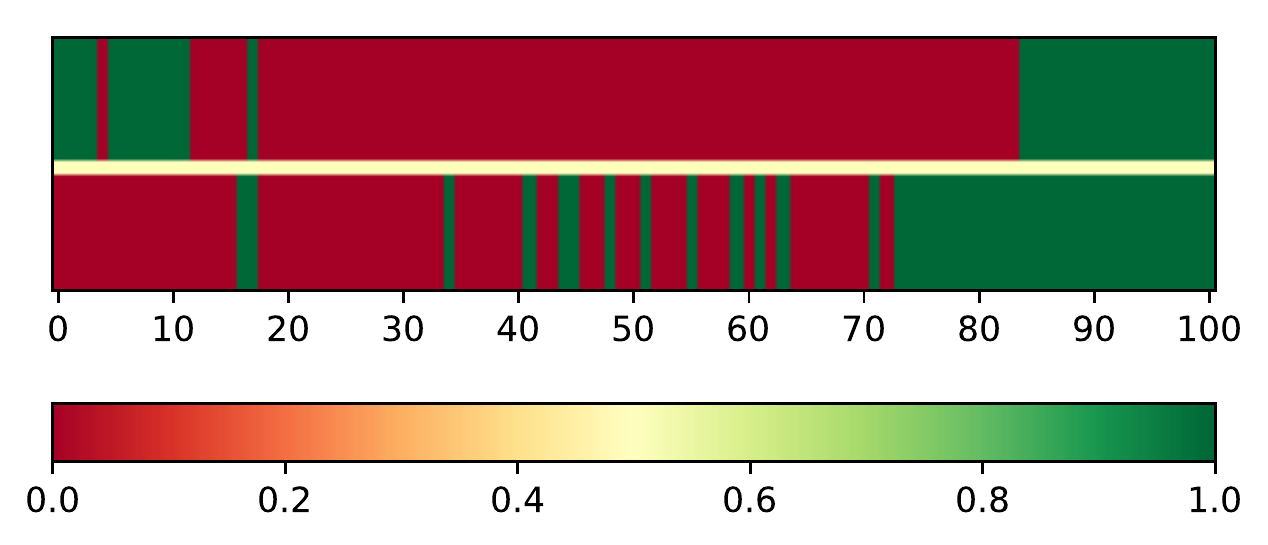}
 \caption{ \label{VN_evolve_strategy_B4_P2} 
 Von Neumann Player ({\bf top}) and Dealer ({\bf bottom})
 chromosome in subsection \ref{results_evolutionary_von_neumann},
 calculated as the final
  average of $\alpha=0.1$ of the most fit Players/Dealers in the population of $5000$ Players/Dealers, after $1000$ rounds of evolution,
 with $10000$ randomly paired games per Player/Dealer on each round. Players ante $a=1$ and bet $b=4$.
 Analytical prediction for the continuous game is $x_1=10$, $x_2=85$, $c=15$,  $y_0=70$.
 }
 \end{center}
\end{figure}

\begin{figure}
\begin{center}
 \includegraphics[width=12cm,height=12cm,keepaspectratio]{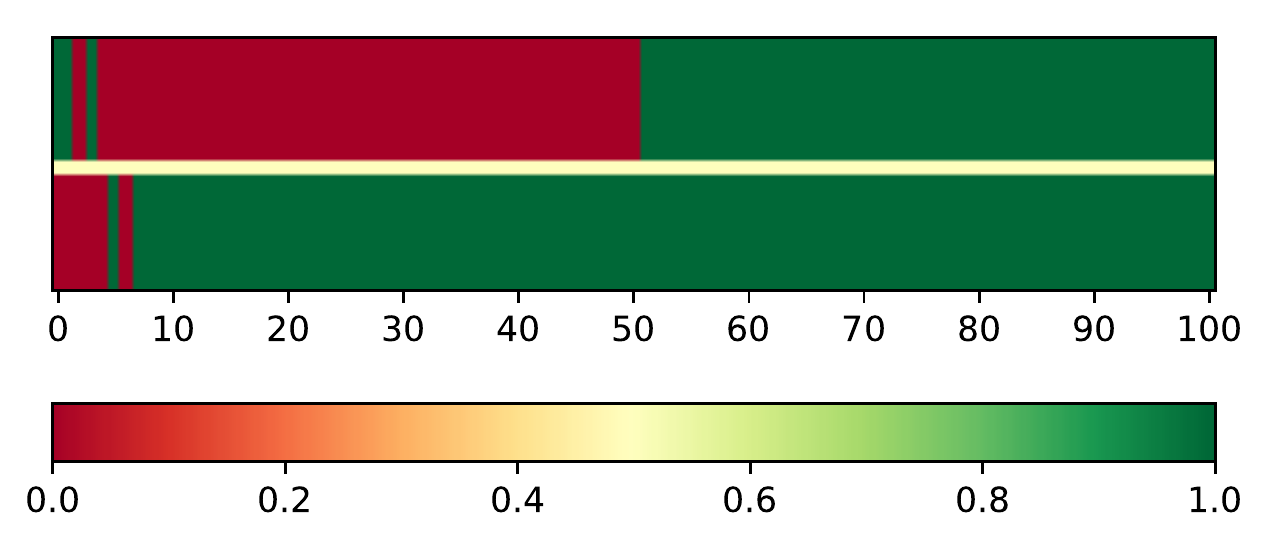}
 \caption{ \label{VN_evolve_strategy_B1_P16} 
 Von Neumann Player ({\bf top}) and Dealer ({\bf bottom})
 chromosome in subsection \ref{results_evolutionary_von_neumann},
 calculated as the final
  average of $\alpha=0.1$ of the most fit Players/Dealers in the population of $5000$ Players/Dealers, after $1000$ rounds of evolution,
 with $10000$ randomly paired games per Player/Dealer on each round. Players ante $a=8$ and bet $b=1$.
 Analytical prediction for the continuous game is $x_1=3$, $x_2=54$, $c=46$,  $y_0=8$.
 }
 \end{center}
\end{figure}

\begin{figure}
\begin{center}
 \includegraphics[width=12cm,height=12cm,keepaspectratio]{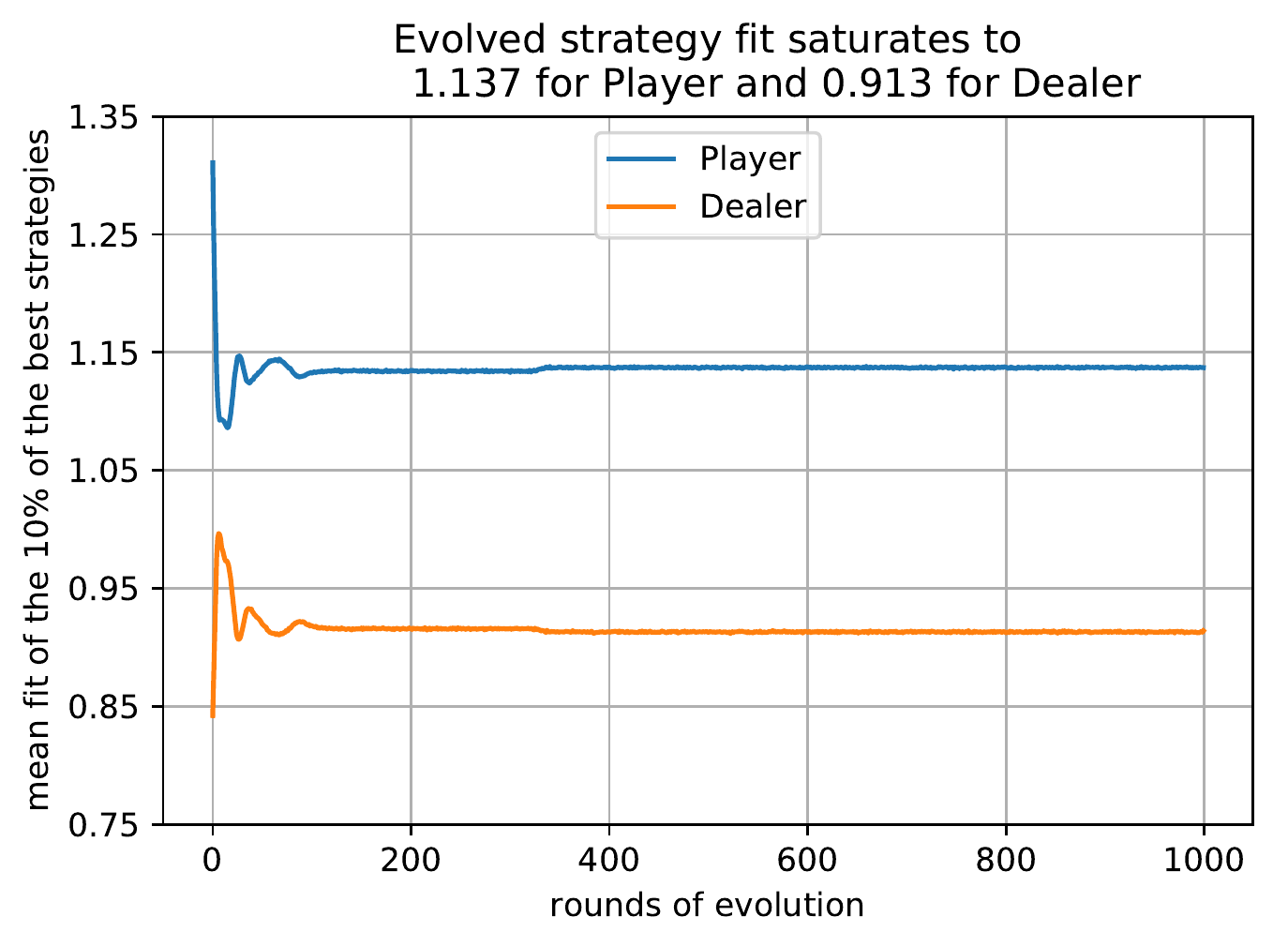}
 \caption{ \label{von_neumann_evolved_player_dealer_fit_B2_P2} 
Player's and Dealer's fit time series (for $\alpha=0.1$ of the most fit) for evolution in subsection \ref{results_evolutionary_von_neumann},
where ante and bet are $(a,b)=(1,2)$.
 }
 \end{center}
\end{figure}

\begin{figure}
\begin{center}
 \includegraphics[width=12cm,height=12cm,keepaspectratio]{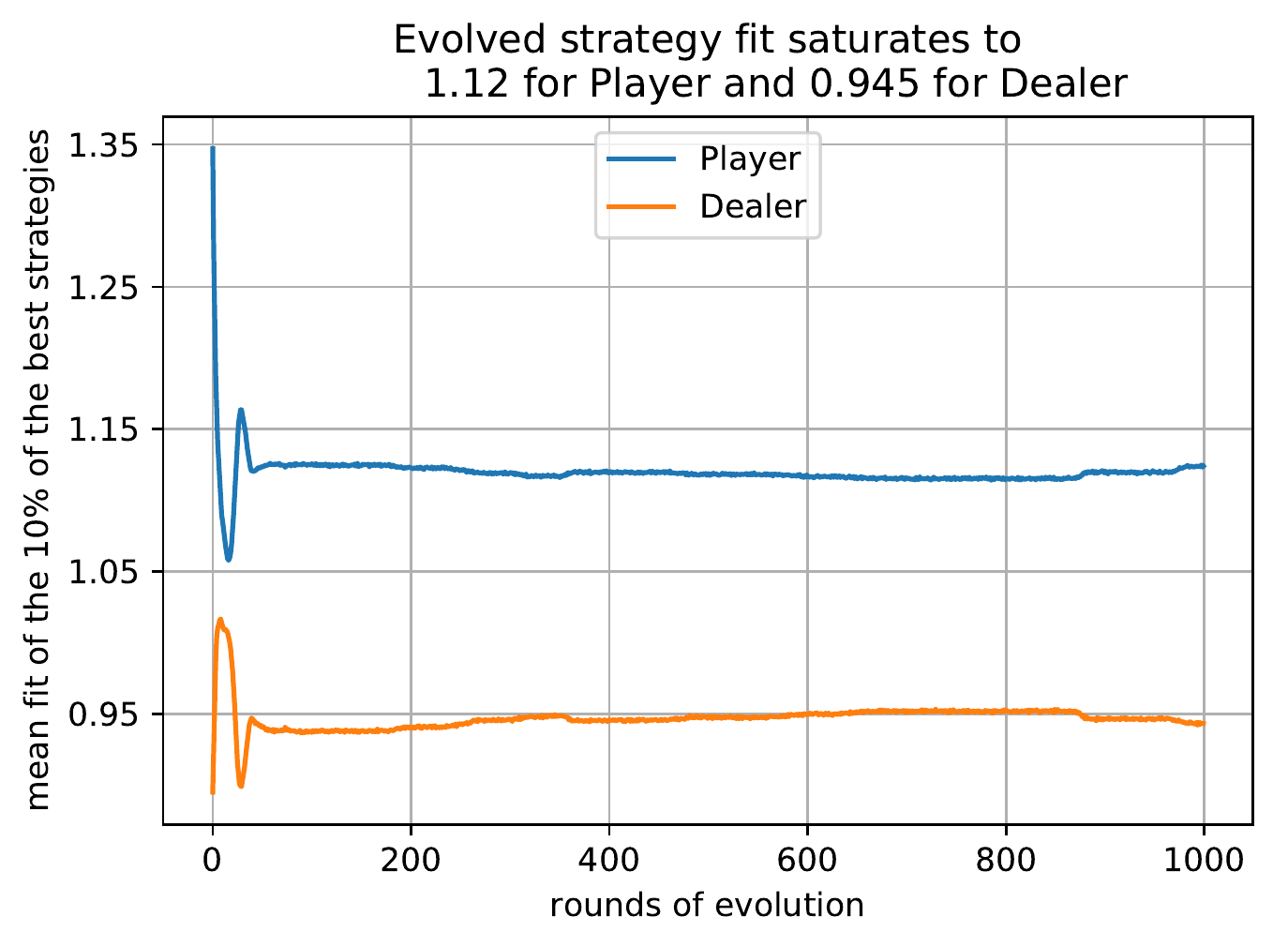}
 \caption{ \label{von_neumann_evolved_player_dealer_fit_B4_P2} 
Player's and Dealer's fit time series (for $\alpha=0.1$ of the most fit) for evolution in subsection \ref{results_evolutionary_von_neumann},
where ante and bet are $(a,b)=(1,4)$.
 }
 \end{center}
\end{figure}

\begin{figure}
\begin{center}
 \includegraphics[width=12cm,height=12cm,keepaspectratio]{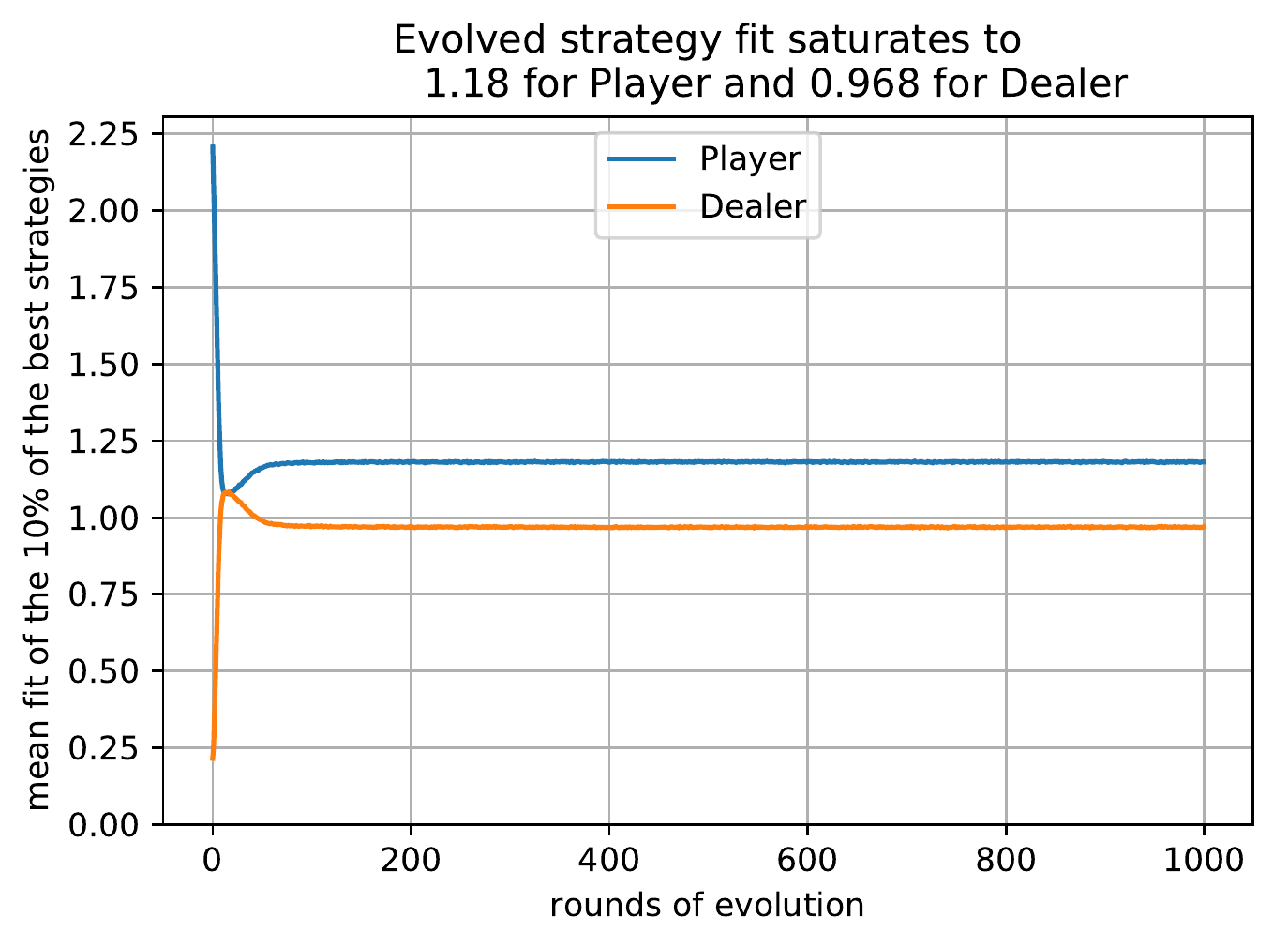}
 \caption{ \label{von_neumann_evolved_player_dealer_fit_B1_P16} 
Player's and Dealer's fit time series (for $\alpha=0.1$ of the most fit) for evolution in subsection \ref{results_evolutionary_von_neumann},
where ante and bet are $(a,b)=(8,1)$.
 }
 \end{center}
\end{figure}

In this subsection we describe our results of applying the genetic algorithm described in
subsection \ref{review_evolutionary_programming} to calculate near-equilibrium strategies 
in the discrete version of the von Neumann poker. The players receive one of $M=100$ numbers,
uniformly spaced in $[0,1)$. Therefore dimension of the Player's and the Dealer's 
chromosome is equal to $M=100$. We will consider games with ante, bet
values $(a,B)$ being $(1,2)$, $(1,4)$, $(8,1)$.

We will consider evolution of the population of $N=5000$ Players and $N=5000$ Dealers,
initialized randomly. At each round of evolution the Players and the Dealers will be paired
up randomly $R=10^4$ times, and play one round of the von Neumann poker after each pairing.
This way each Player and each Dealer will be able to apply its strategy against the average
opponent in the population. At the beginning of each round of evolution all the members of
the Player and Dealer populations have their bankrolls reset to $B_0=10^4$,
a value chosen to be sufficiently large so that no player ends up with a negative bankroll during the game. During $R$
rounds of play the profits and losses accumulate, and at the end the Players and the Dealers
are ranked by their fit scores, $\phi=B_R/B_0$, determined by the final bankroll $B_R$.

The fraction $\alpha=0.1$ of the most fit Players/Dealers are selected for reproduction. The parents of Players and Dealers
are then selected in pairs randomly, with the probabilities
proportional to their fit scores, until they replenish the population to $N=5000$ Players/Dealers.
When two parents produce an offspring, if the parents have the same gene at the
given chromosome slot, then the child will have the same gene with the probability $1-\pi$,
where $\pi=10^{-6}$ is the probability of mutation. 

After $T=1000$ rounds of evolution the $\alpha=0.1$ of the most fit Players/Dealers
are selected, and the population average chromosomes of the Players and Dealers
are calculated. We present the resulting chromosomes for the Player and the Dealer in figure
\ref{VN_evolve_strategy_B2_P2},
for $(a,B)=(1,2)$,
figure \ref{VN_evolve_strategy_B4_P2} for $(a,B)=(1,4)$, and
figure \ref{VN_evolve_strategy_B1_P16}
for $(a,B)=(8,1)$. We also plot the evolution
time series of the mean fit scores $B_R/B_0$ of the $\alpha=0.1$ of the most fit Players and Dealers
in figures \ref{von_neumann_evolved_player_dealer_fit_B2_P2},
\ref{von_neumann_evolved_player_dealer_fit_B4_P2}, and
\ref{von_neumann_evolved_player_dealer_fit_B1_P16}. Notice that at some points it looks
like the Player's and Dealer's payoffs do not sum up to zero. This is because we used the highest performing
Players/Dealers to calculate the mean fit scores, so the players who appeared
on the graphs did not necessarily win the money from each other.

\section{Counterfactual regret minimization in the von Neumann poker}
\label{CFR_von_neumann}

In this section we are going to describe our results of application of the
counterfactual regret minimization algorithm (typically abbreviated as CFR) \cite{Zinkevich2007}
to calculation of the near-equilibrium strategies in the von Neumann poker.
We refer the reader to \cite{Neller2013} for an excellent review of the
counterfactual regret minimization algorithm, as well as
the regret matching algorithm \cite{Hart2000}. In subsection 
\ref{CFR_review} we outline the principles of the CFR algorithm on the example of the von Neumann poker,
and in subsection \ref{CFR_von_neumann_results}
we describe our results.
Algorithm discussed in subsection \ref{CFR_review} is also applicable, with minor adjustments,
to section \ref{sec:flop_poker_cfr} where the CFR is used to compute near-Nash equilibrium in the flop poker.

In subsection \ref{CFR_von_neumann_results} we show that the CFR algorithm in general finds an
optimal equilibrium strategy of the von Neumann poker which is not admissible,
in the sense defined in subsection \ref{von_neumann_solution}. This is to be contrasted
with the genetic algorithm yielding a close to admissible output,
as described in subsection \ref{results_evolutionary_von_neumann}. Therefore at least
in its simplest versions the CFR algorithm finds one equilibrium strategy at random,
of the many equilibrium strategies which might exist.

\subsection{CFR algorithm}
\label{CFR_review}

Counterfactual regret minimization algorithm can be used to calculate the Nash equilibrium
(or rather the  $\epsilon$-Nash equilibrium, which takes into account the convergence 
speed of the algorithm, and puts a bound on how close one gets to the true Nash equilibrium)
solution to the two-person zero-sum games with incomplete information \cite{Zinkevich2007}.
The CFR algorithm and its improvements have been at the core of building the most recent
top poker playing agents, such as \cite{Bowling2015,Moravc2017,Brown2017}.
In this subsection we are going to describe implementation of the CFR algorithm of \cite{Zinkevich2007} to find
(approximate) Nash equilibrium strategies in the von Neumann poker.  

Similar to the genetic algorithm described in section \ref{sec:von_neumann_evolve},
the CFR algorithm aims to learn the Nash equilibrium through the self-play. Unlike the genetic algorithm,
the CFR does not require initializing the entire population of Players and Dealers. Instead, just one
Player and just one Dealer are initialized, and play against each other. Each Player is attributed with the
vector $V_P$ which defines its current strategy, the vector $S_P$ which stores its cumulative
strategy, and the vector $R_P$, which stores its cumulative regret (to be explained below).
Similarly, each Dealer possesses the instantaneous strategy vector $V_D$, the cumulative strategy vector $S_D$,
and the cumulative regret vector $R_D$. Each of these vectors,
$V_{P,D}$, $S_{P,D}$, $R_{P,D}$, 
has the length $M$ (for the von Neumann poker $M=100$, for the flop poker, defined in 
section \ref{sec:flop_poker_cfr}, $M=169$), equal to the number of possible hands which the Player/Dealer can be dealt in the game.
Each entry of the vectors $S_{P,D}$, $R_{P,D}$ is itself a vector of size 2, corresponding to two pure strategies
which can be played by the Player (bet/check) and the Dealer (call/fold) at each possible decision node
of the game tree.

The entry $V_P(i)$ of the Player's current strategy vector $V_P$ prescribes the probability with
which the Player will bet when holding the hand $i$. Similarly, the entry $V_D(j)$ of the Dealer's
current strategy vector $V_D$ prescribes the probability with which the Dealer will call (if facing bet) when dealt
the hand $j$.
At the beginning of the algorithm all the entries of the current
strategy vectors $V_P(i)$, $V_D(j)$, $i,j=1,\dots,100$, are initialized to $0.5$.
All the entries of the cumulative vectors $S_{P,D}$, $R_{P,D}$ are initialized to zero.

The following training through self-play then takes place. The Player and the Dealer play for $T$ rounds.
At each round $t=1,\dots,T$ the Player and the Dealer play one game of von Neumann poker.
In the given round denote the hand received by the Player as $i$, and the hand received by the Dealer as $j$.
The CFR algorithm calculates the regret of
not playing each of the pure strategies (bet and check for the Player, and call and fold for the Dealer) rather than
using the current (mixed) strategies $V_P(i)$, $V_D(j)$.
The counterfactual aspect of it (as opposed to factually playing with the strategies $V_P(i)$, $V_D(j)$)
consists of iterating over all possible pure strategies, and comparing the outcome of playing those pure strategies
($V_P^{{\rm bet},{\rm check}}(i)=1,0$ for Player, and $V_D^{{\rm call},{\rm fold}}(j)=1,0$ for Dealer)
which would have happened, with the outcome of playing the given strategies $V_P(i)$, $V_D(j)$
which did happen.

In other words, the regret of not using each pure strategy is calculated as a difference
between the expected value of using the current strategy, $E_P(i)$, $E_D(j)$,
and the expected value of using
the pure strategies, $E_P^{{\rm bet}, {\rm check}}(i)$, $E_D^{{\rm call},{\rm fold}}(j)$.
These expected values depend on what game state the player is in, that is, whether $i=j$ (draw),
$i>j$ (Player wins), or $i<j$ (Dealer wins). Let us consider each of these possible game states
separately. (Unlike the pot framework used in section \ref{von_neumann}, here we are going to use
the zero-sum framework. Denote $a$ to be the ante, and $B$ to be the bet.)
\begin{itemize}
\item
{\bf Draw,} $i=j$ 
\begin{align}
E_P(i)&=V_P(i)(1-V_D(j))a \\ & {\rm counterfactual}\quad=\quad  \left \{
  \begin{aligned}
    &E_P^{{\rm bet}}(i)=(1-V_D(j))a \\
    &E_P^{{\rm check}}(i)=0\,
  \end{aligned} \right.\\
E_D(j)&=-(1-V_D(j))a  \\ & {\rm counterfactual}\quad =\quad  \left \{
  \begin{aligned}
    &E_D^{{\rm call}}(j)=0 \\
    &E_D^{{\rm fold}}(j)=-a\,
  \end{aligned} \right.
\end{align}
\item
{\bf Player wins,} $i>j$ 
\begin{align}
E_P(i)&=V_P(i)(V_D(j)(a+B)+(1-V_D(j))a)+(1-V_P(i))a \\ & {\rm counterfactual}\quad =\quad  \left \{
  \begin{aligned}
    &E_P^{{\rm bet}}(i)=V_D(j)(a+B)+(1-V_D(j))a \\
    &E_P^{{\rm check}}(i)=a\,
  \end{aligned} \right.\\
E_D(j)&=-V_D(j)(a+b)-(1-V_D(j))a \\ & {\rm counterfactual}\quad =\quad  \left \{
  \begin{aligned}
    &E_D^{{\rm call}}(j)=-a-B \\
    &E_D^{{\rm fold}}(j)=-a\,
  \end{aligned} \right.
\end{align}
\item
{\bf Dealer wins,} $i<j$ 
\begin{align}
E_P(i)&=V_P(i)(-V_D(j)(a+B)+(1-V_D(j))a)-(1-V_P(i))a \\ & {\rm counterfactual}\quad = \quad   \left \{
  \begin{aligned}
    &E_P^{{\rm bet}}(i)=-V_D(j)(a+B)+(1-V_D(j))a \\
    &E_P^{{\rm check}}(i)=-a\,
  \end{aligned} \right.\\ 
E_D(j)&=V_D(j)(a+b)-(1-V_D(j))a \\  & {\rm counterfactual}\quad = \quad \left \{
  \begin{aligned}
    &E_D^{{\rm call}}(j)=a+B \\
    &E_D^{{\rm fold}}(j)=-a\,
  \end{aligned} \right.
\end{align}
\end{itemize}
The counterfactual regrets are then calculated,
\begin{align}
\Delta R_P(i)&=\left \{
  \begin{aligned}
    &\Delta R_P^{{\rm bet}}(i)=E_P^{{\rm bet}}(i)-E_P(i) \\
    &\Delta R_P^{{\rm check}}(i)=E_P^{{\rm check}}(i)-E_P(i)\,
  \end{aligned} \right.\\
  \Delta R_D(j)&=\left \{
  \begin{aligned}
    &\Delta R_D^{{\rm call}}(j)=V_P(i)(E_D^{{\rm call}}(j)-E_D(j)) \\
    &\Delta R_D^{{\rm fold}}(j)=V_P(i)(E_D^{{\rm fold}}(j)-E_D(j))\,
  \end{aligned} \right. 
\end{align}
where regrets of the Dealer are weighted by the probability $V_P(i)$ (of the Player placing the bet) to get to the state where the Dealer
is faced with the decision to call or fold.

The regrets $\left(\Delta R_P^{{\rm bet}}(i), \Delta R_P^{{\rm check}}(i)\right)$,
$\left(\Delta R_D^{{\rm call}}(j),\Delta R_D^{{\rm fold}}(j)\right)$ are then added to the  cumulative regret vectors components
$R_P(i)$, $R_D(j)$. 
The negative components of $R_P(i)$ or $R_D(j)$, if exist, are replaced with zeros.
The current strategy vectors components $V_P(i)$, $V_D(j)$ are subsequently updated. 
If both of the entries of $R_P(i)$ are zero then $V_P(i)$ is set to $0.5$. Else, we set
\begin{equation}
\label{SP_strategy_update}
V_P(i)=\frac{R_P^{{\rm bet}}(i)}{R_P^{{\rm bet}}(i)+R_P^{{\rm check}}(i)}\,,
\end{equation}
and similarly for the Dealer's
strategy $V_D(j)$. Finally, the cumulative strategy vectors $S_P(i)$, $S_D(j)$ are incremented by
$\left(V_P^{{\rm bet}}(i),1-V_P^{{\rm bet}}(i)\right)$, $\left(V_D^{{\rm call}}(j),1-V_D^{{\rm call}}(j)\right)$, respectively.

At the end of the training the cumulative strategy vectors are used to calculate the final output strategies
(for all $i,j=1,\dots,M$),
\begin{align}
W_P(i)&=\frac{S_P^{{\rm bet}}(i)}{S_P^{{\rm bet}}(i)+S_P^{{\rm check}}(i)}\,,\\
W_D(j)&=\frac{S_D^{{\rm call}}(j)}{S_D^{{\rm call}}(j)+S_D^{{\rm call}}(j)}\,.
\end{align}
The statement is that $W_P(i)$ converges to the Nash equilibrium for the Player's probability to bet
with the hand $i$, and $W_D(j)$ converges to the Nash equilibrium for the Dealer's probability
to call with the hand $j$ \cite{Zinkevich2007}.

\subsection{Results of the CFR calculation of the von Neumann poker strategies}
\label{CFR_von_neumann_results}

\begin{figure}
\begin{center}
 \includegraphics[width=12cm,height=12cm,keepaspectratio]{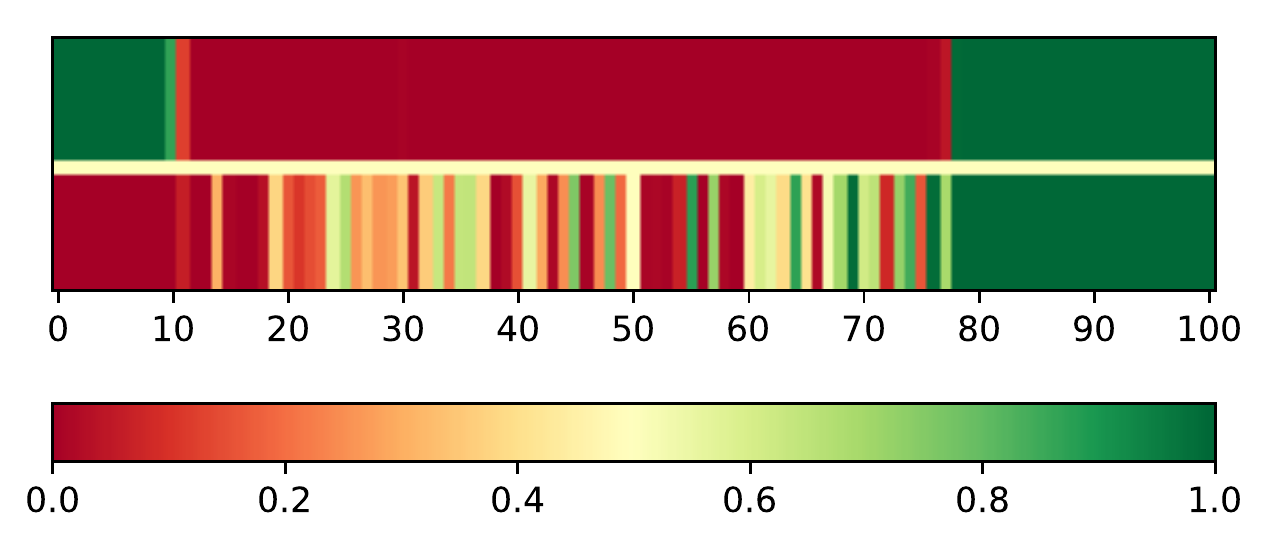}
 \caption{ \label{VN_regret_strategy_B2_P2} 
 Von Neumann Player strategy ({\bf top}) and Dealer strategy ({\bf bottom}) in subsection \ref{CFR_von_neumann_results},
 calculated as the final
  output after $10^{10}$ rounds of self-play and training using the CFR algorithm. Players ante $a=1$ and bet $b=2$.
 Analytical prediction for the continuous game is $x_1=11$, $x_2=78$, $c=22$, $y_0=56$.
 Notice that while
 the found Dealer's strategy is not admissible, it satisfies the equilibrium constraint by exhibiting $c\simeq 22.9$.
 }
 \end{center}
\end{figure}

\begin{figure}
\begin{center}
 \includegraphics[width=12cm,height=12cm,keepaspectratio]{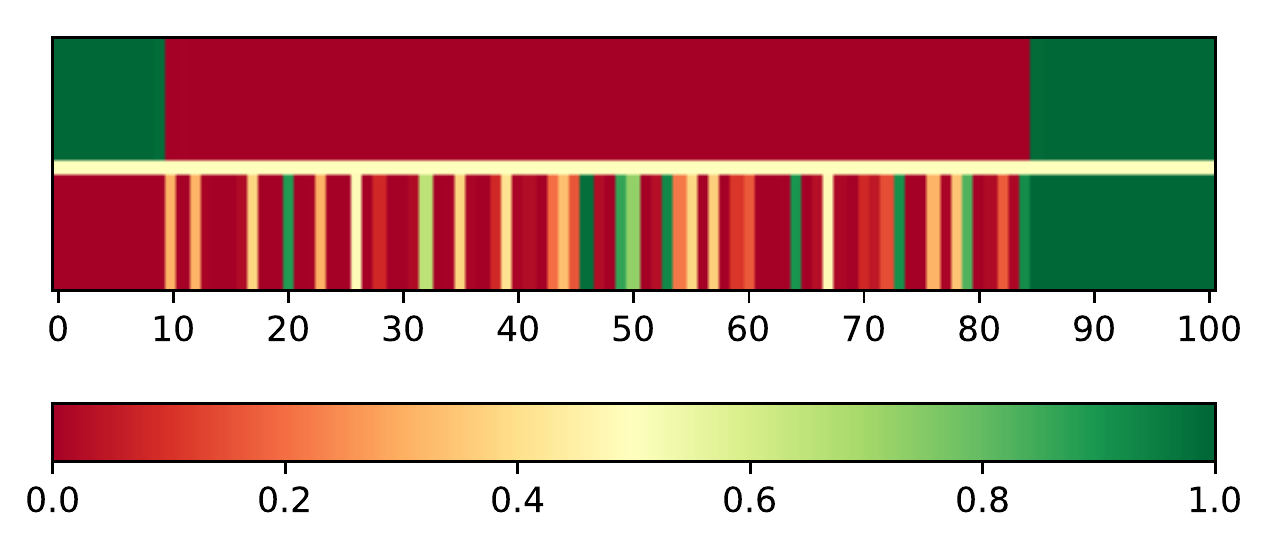}
 \caption{ \label{VN_regret_strategy_B4_P2} 
 Von Neumann Player strategy ({\bf top}) and Dealer strategy ({\bf bottom}) in subsection \ref{CFR_von_neumann_results},
 calculated as the final
  output after $10^{10}$ rounds of self-play and training using the CFR algorithm. Players ante $a=1$ and bet $b=4$.
 Analytical prediction for the continuous game is $x_1=10$, $x_2=85$, $c=15$, $y_0=70$.
 Notice that while
 the found Dealer's strategy is not admissible, it satisfies the equilibrium constraint by exhibiting $c\simeq 15.3$.
 }
 \end{center}
\end{figure}

\begin{figure}
\begin{center}
 \includegraphics[width=12cm,height=12cm,keepaspectratio]{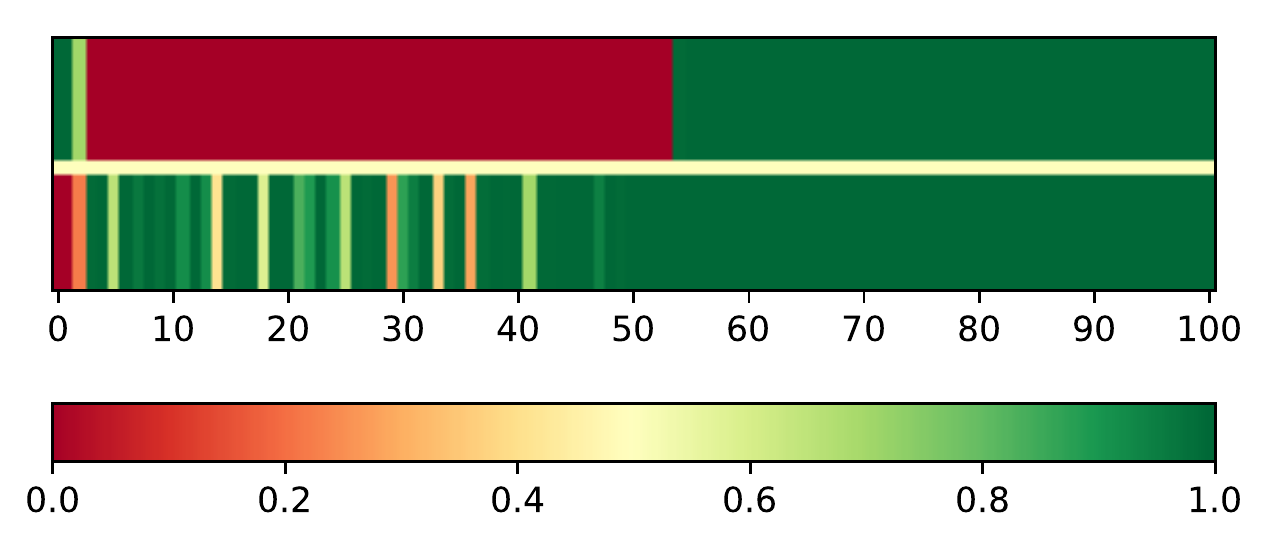}
 \caption{ \label{VN_regret_strategy_B1_P16} 
 Von Neumann Player strategy ({\bf top}) and Dealer strategy ({\bf bottom}) in subsection \ref{CFR_von_neumann_results},
 calculated as the final
  output after $10^{10}$ rounds of self-play and training using the CFR algorithm. Players ante $a=8$ and bet $b=1$.
 Analytical prediction for the continuous game is $x_1=3$, $x_2=54$, $c=46$, $y_0=8$.
 Notice that while
 the found Dealer's strategy is not admissible, it satisfies the equilibrium constraint by exhibiting $c\simeq 46.2$.
 }
 \end{center}
\end{figure}

In this subsection we provide the results of applying the CFR algorithm to find the near-equilibrium strategies in the von Neumann poker with
$(a,B)=(1,2)$, see figure \ref{VN_regret_strategy_B2_P2},
$(a,B)=(1,4)$, see figure \ref{VN_regret_strategy_B4_P2},  and
$(a,B)=(8,1)$, see figure \ref{VN_regret_strategy_B1_P16}.
We run the training over $10^{10}$ steps, but the strategy converges to equilibrium
much sooner (the output after $10^8$ steps of self-training already shows the equilibrium strategy rather
accurately).
Notice that the CFR algorithm finds the correct (equilibrium) Player's and Dealer's strategies,
in agreement with the analytical results, reviewed in subsection \ref{von_neumann_solution}.
As it was pointed out in subsection \ref{von_neumann_solution} there are infinitely
many equilibrium Dealer's strategies. All these strategies satisfy the constraint 
(\ref{c_def}), (\ref{c_solution}). Apart from this constraint the Dealer's probability $q(y)$ in $(x_1,x_2)$
can be arbitrary (as long as the corresponding $e_1(x)<0$ in $(x_1,x_2)$). From our results we
see that indeed in every case $q(y)=0$ for $y\in [0,x_1)$, $q(y)=1$ for $y\in (x_2,1]$, and $\int _{x_1}^{x_2}
dy\, q(y)\simeq c$, where $c$ is given by (\ref{c_solution}). Other than that the results of our CFR calculations show
that the specific $q(y)$, $y\in (x_1,x_2)$ are arbitrary, as long as the constraints mentioned above 
are satisfied. That is, the CFR algorithm finds the Dealer's strategy which is equilibrium,
but not necessarily admissible, as defined in subsection \ref{von_neumann_solution}.
This is to be contrasted with the output of the evolutionary optimization described
in subsection \ref{results_evolutionary_von_neumann}, which approximates well the strategy
which is both equilibrium and admissible.

\section{Flop poker}
\label{sec:flop_poker}

In this section we consider the game of flop poker, which can be seen as a simplified
version of the Texas Hold'Em, and as a natural
upgrade of the von Neumann poker in the direction of the real poker games.
We begin by describing the rules of the game.
In the flop poker two players are in a heads up game.
Before the round of game each player puts an ante $a$ into the pot. Each player is dealt two
private cards out of 52-card deck. Then the first player (Player) can choose to bet $b$ or check.
If the Player bets, then the second player (Dealer) can either call the bet $b$, or fold. If the Dealer
folds, then the Player collects the entire pot. If the Dealer calls (or if the Player checks),
then three cards are dealt on the
table (community cards), and the player who makes the highest five-card hand (composed by two private cards
and three community cards) wins.
Versions of the flop poker game exist in the literature, in particular in \cite{Barone1999},
which discussed the game in which five community cards are dealt, and each of the two players can
make the best five-card hand out of two private cards and five community cards (see also \cite{Selby1999}).

Due to non-uniform
probability distributions of getting various poker hands (as discussed in subsection 
\ref{introducing_von_neumann}) this game is less tractable analytically
than the von Neumann poker (versions of the game can be solved using the linear
programming methods \cite{Selby1999}).
Besides, unlike the von Neumann poker, in the flop poker (just as in the real poker games) any private
hand can end up making the strongest final hand, given the appropriate community cards. 
In this section we are going to derive expressions
defining the Nash equilibrium strategies of the flop poker players.
In section \ref{sec:flop_poker_evolution} we will apply the genetic algorithm
to calculate the near-equilibrium flop poker strategies, and test its output against the theoretical
predictions given in this section. In section \ref{sec:flop_poker_cfr} we will apply the counterfactual regret minimization
algorithm to calculate the $\epsilon$-Nash equilibrium in the flop poker.

Derivation
of expressions for the Nash equilibrium in the flop poker follows the similar calculation for the von Neumann poker, given in subsection \ref{von_neumann_solution}.
We denote $P=2a$ to be the pot composed of the initial antes $a$ of the players. 
We will be working in the pot framework, so that the
game is represented as $P$-sum, rather than zero-sum (as discussed in subsection \ref{von_neumann_solution}
this is just a matter of convenience). To make it a zero sum
game we should subtract $P/2$ ante from the expected winning of each player.

We will be using $p(i)$ to denote the probability that the Player will bet when
holding the hand $i$, and $q(j)$ to denote the probability that the Dealer will
call (if facing bet) when holding the hand $j$.
Each private hand (a pair of two cards held by each player) can take one of 169 values:
13 pairs, ${13\choose 2}=78$ suited non-pairs, and ${13\choose 2}=78$ non-suited non-pairs.
Notice that when the suit degeneracy is not taken into account, the total number
of ways to deal two cards out of 52-card deck is 1326, however these hands can be grouped into
only 169 distinct categories, where degeneracy of each hand is 6 in the pairs category, 4 in the suited non-pair
category, and 12 in the non-suited non-pair category.
The goal is to derive Nash equilibrium values 
for the players strategies $\{p(i)\}$ and $\{q(j)\}$, $i,j=1,\dots,169$. 

Denote $h(i)$ to be the probability to receive the hand $i$, see (\ref{h_of_i_def}).
Denote $h(j|i)$ to be the conditional probability that a player has the hand $j$,
given that it is known that their opponent has the hand $i$. Denote $w(i|j)$
to be the conditional probability to win with the hand $i$, given that the opponent has the
hand $j$, where to win, in this case, means to make a better five-card hand after the flop.
Similarly, denote $d(i|j)$ to be the conditional probability to draw with the hand $i$
against the opponent's $j$. Clearly $d(i|j)=d(j|i)$, and $w(i|j)+w(j|i)+d(i|j)=1$. 

Expected value of the winnings of the Player holding the hand $i$ is given by
\begin{align}
\label{EP_flop}
E_P(i)=p(i)e_P(i)+P\,\left(W(i)+\frac{1}{2}\,D(i)\right)\,,
\end{align}
where we denoted the probabilities to win and draw with the hand $i$ as 
(here and below sums over $i$, $j$ stand for sums over 169 possible private hands of two cards)
\begin{equation}
\label{flop_probabilities_to_win_and_draw_}
W(i)=\sum _jh(j|i)w(i|j)\,,\qquad D(i)=\sum _jh(j|i)d(i|j)\,,
\end{equation}
and introduced
\begin{equation}
\label{eP_def}
e_P(i)=\sum _j h(j|i)\left[(P+B)w(i|j)-Bw(j|i)-P+\frac{P}{2}\,d(i|j)\right]q(j)+P\left(1-W(i)-\frac{1}{2}\,D(i)\right)\,.
\end{equation}

Similarly, the expected value of the Dealer's winnings when holding the hand $j$ is determined by
\begin{align}
\label{ED_flop}
E_D(j)=q(j)e_D(j)+P\,\left(W(j)+\frac{1}{2}\,D(j)-\sum _i h(i|j)\left(w(j|i)+\frac{1}{2}d(j|i)\right)p(i)\right)\,,
\end{align}
where we introduced
\begin{equation}
\label{eD_def}
e_D(j)=\sum _i h(i|j)\left[(P+B)w(j|i)-Bw(i|j)+\frac{P}{2}\,d(j|i)\right]p(i)\,.
\end{equation}
The probabilities $w(i|j)$, $d(i|j)$ can be obtained by simulation.

From the expressions (\ref{EP_flop}), (\ref{ED_flop}) it follows that the 
Nash equilibrium strategies for the Player will be such that $p(i)=1$
when $e_P(i)>0$, and $p(i)=0$ when $e_P(i)<0$. Similarly, for the
Dealer we obtain $q(j)=1$
when $e_D(j)>0$, and $q(j)=0$ when $e_D(j)<0$. If it happens that $e_P(i)=0$ ($e_D(j)=0$)
then the Player (Dealer) will be indifferent to the choice of the betting (calling) probability with
the hand $i$ ($j$).

Despite the similarity with the analogous expression (\ref{e1_def}),
(\ref{e2_def}) in the von Neumann poker, finding the Nash equilibrium
strategy in the flop poker is less tractable analytically. This is because, unlike the von Neumann poker,
the probabilities $h(i)$, $h(j|i)$
of getting various hands are no longer uniform,
and the probabilities $w(i|j)$, $d(i|j)$ of winning and drawing are not simply determined by the relative values of $i$ and
$j$ to be equal to either zero or one.

In fact in the flop poker just as in the real poker (and unlike the von Neumann
poker)
any hand can win, and therefore even the ranking of the private hands $i$ is non-trivial.
One possibility to rank the hand would be by the probability $W(i)$, defined in (\ref{flop_probabilities_to_win_and_draw_}), to out-flop the opponent (this is
sometimes referred to as ranking by the roll-out simulations, especially when all five community cards are
dealt, and the player who makes the best five-card hand wins). When combined
with such considerations as the number of players, position in the game, and the actions of other players,
such ranking of strength of the poker hands resembles the 
known Sklansky ranking \cite{Sklansky1999} (see also \cite{Billings2002,Davidson2002}).
Notice that the ranking based on the probability to make the best hand depends on
whether we consider the best hand on the flop (in which case the worst hand is 32o), or on the river
(in which case the worst hand is 72o), and the precise numbers also depend on the number of players
\cite{Sklansky1999}.

\section{Evolutionary optimization of the flop poker strategies}
\label{sec:flop_poker_evolution}

\begin{figure}[!tbp]
  \centering
  \begin{minipage}[b]{0.4\textwidth}
     \includegraphics[width=7cm,height=7cm,keepaspectratio]{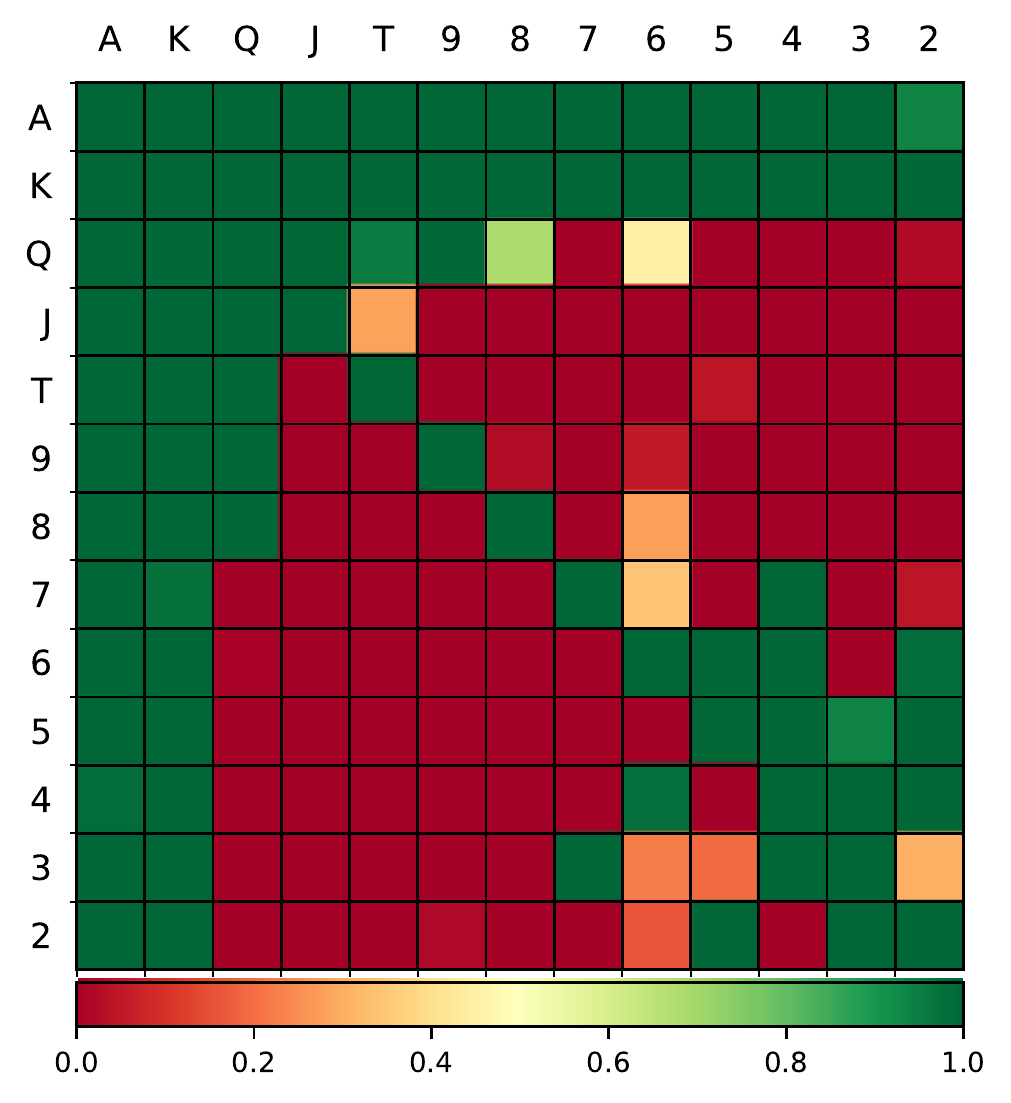}
 \caption{ \label{Flop_evolve_player_strategy_B2_P2} 
 Player's strategy in section \ref{sec:flop_poker_evolution}, evolved for $(a,B)~=~(1,2)$.
 }
  \end{minipage}
  \hfill
  \begin{minipage}[b]{0.4\textwidth}
   \includegraphics[width=7cm,height=7cm,keepaspectratio]{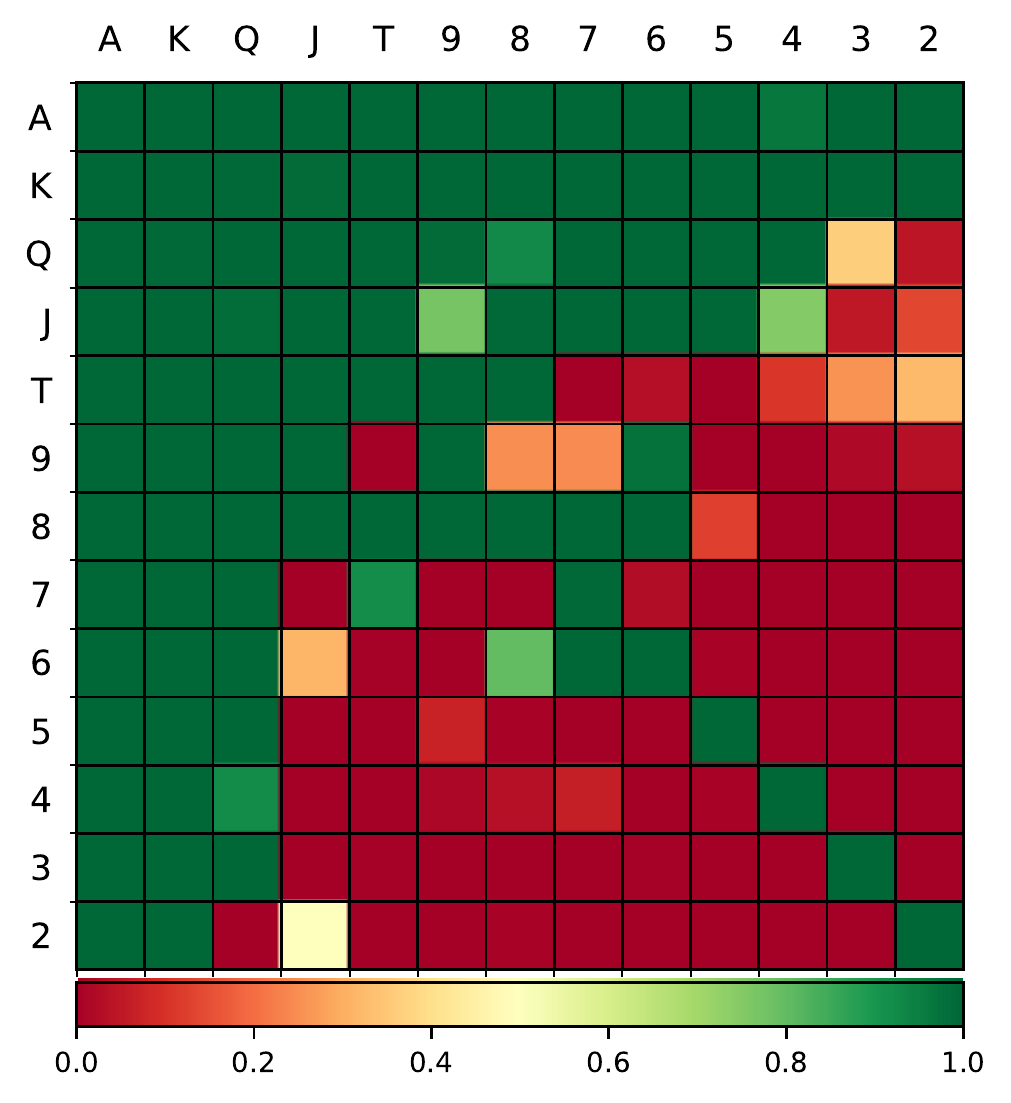}
 \caption{ \label{Flop_evolve_dealer_strategy_B2_P2} 
 Dealer's strategy in section \ref{sec:flop_poker_evolution}, evolved for $(a,B)~=~(1,2)$.
 }
  \end{minipage}
\end{figure}

\begin{figure}[!tbp]
  \centering
  \begin{minipage}[b]{0.4\textwidth}
     \includegraphics[width=7cm,height=7cm,keepaspectratio]{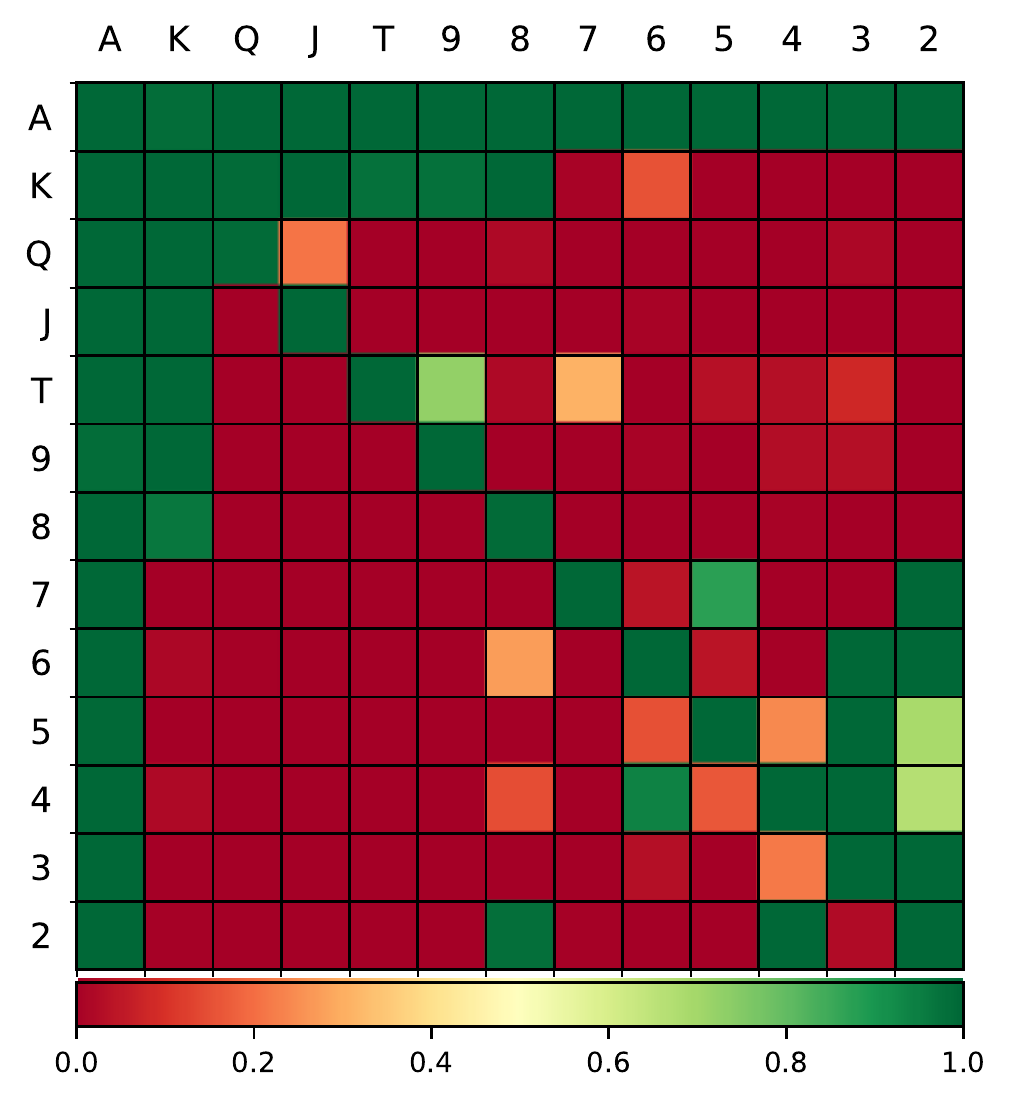}
 \caption{ \label{Flop_evolve_player_strategy_B4_P2} 
 Player's strategy in section \ref{sec:flop_poker_evolution}, evolved for $(a,B)~=~(1,4)$.
 }
  \end{minipage}
  \hfill
  \begin{minipage}[b]{0.4\textwidth}
   \includegraphics[width=7cm,height=7cm,keepaspectratio]{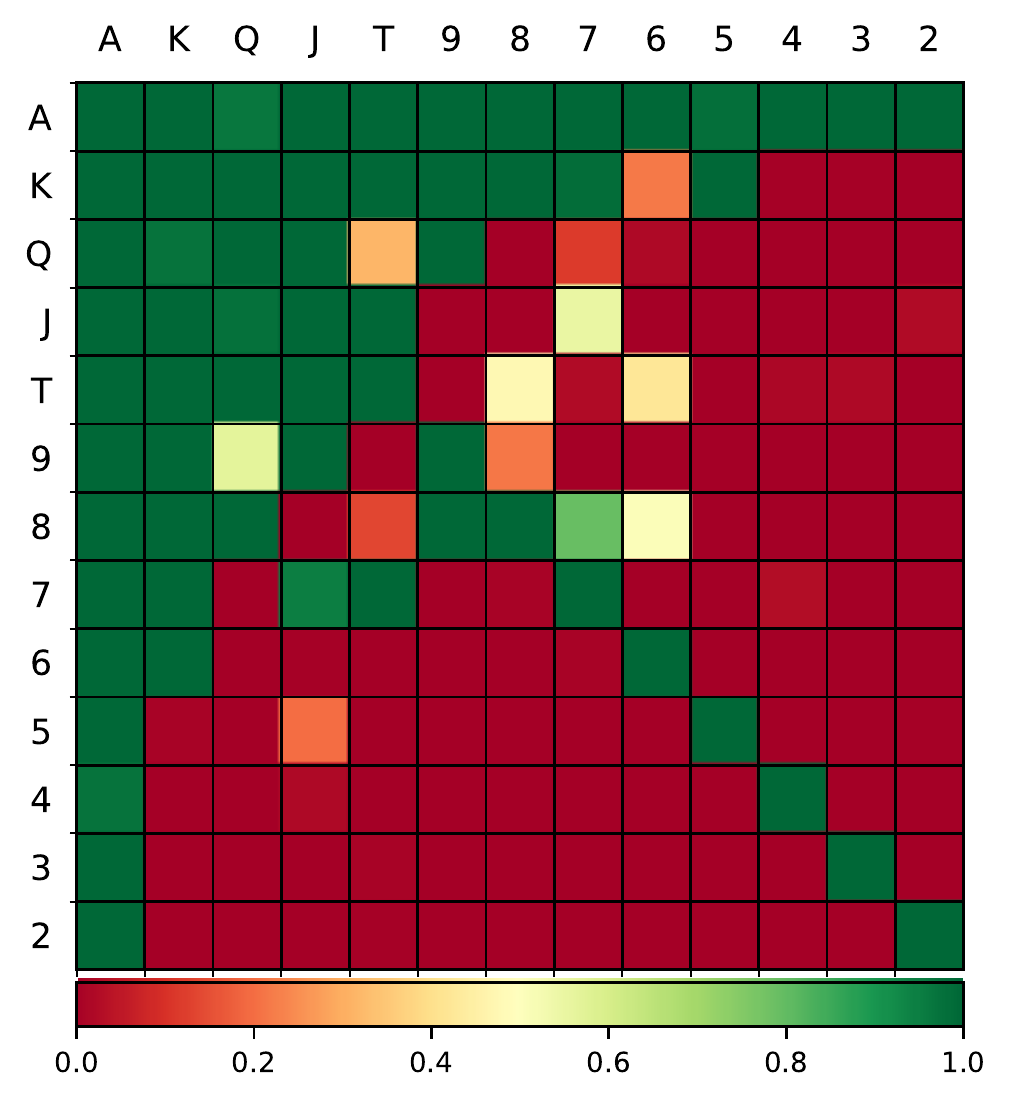}
 \caption{ \label{Flop_evolve_dealer_strategy_B4_P2} 
 Dealer's strategy in section \ref{sec:flop_poker_evolution}, evolved for $(a,B)~=~(1,4)$.
 }
  \end{minipage}
\end{figure}

\begin{figure}[!tbp]
  \centering
  \begin{minipage}[b]{0.4\textwidth}
     \includegraphics[width=7cm,height=7cm,keepaspectratio]{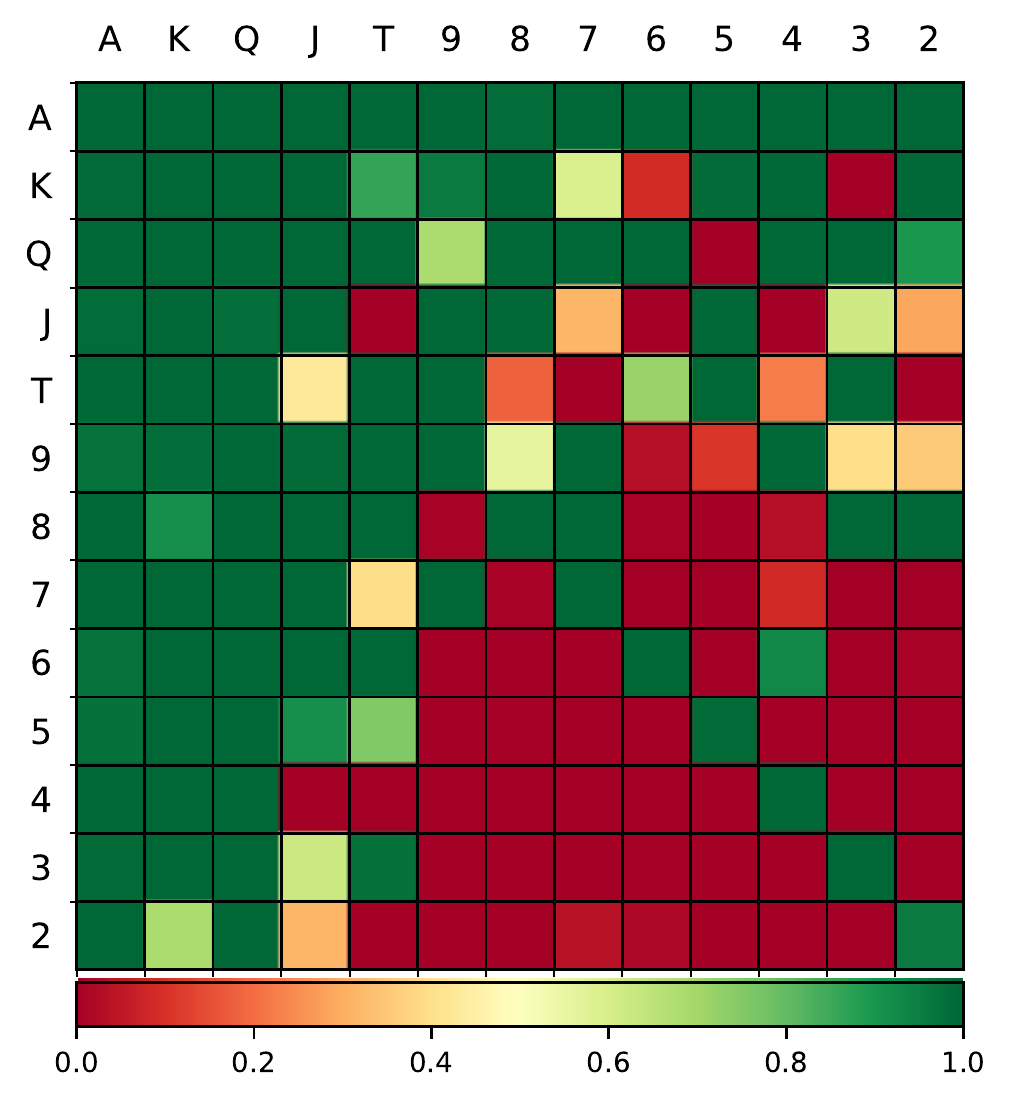}
 \caption{ \label{Flop_evolve_player_strategy_B1_P16} 
 Player's strategy in section \ref{sec:flop_poker_evolution}, evolved for $(a,B)~=~(8,1)$.
 }
  \end{minipage}
  \hfill
  \begin{minipage}[b]{0.4\textwidth}
   \includegraphics[width=7cm,height=7cm,keepaspectratio]{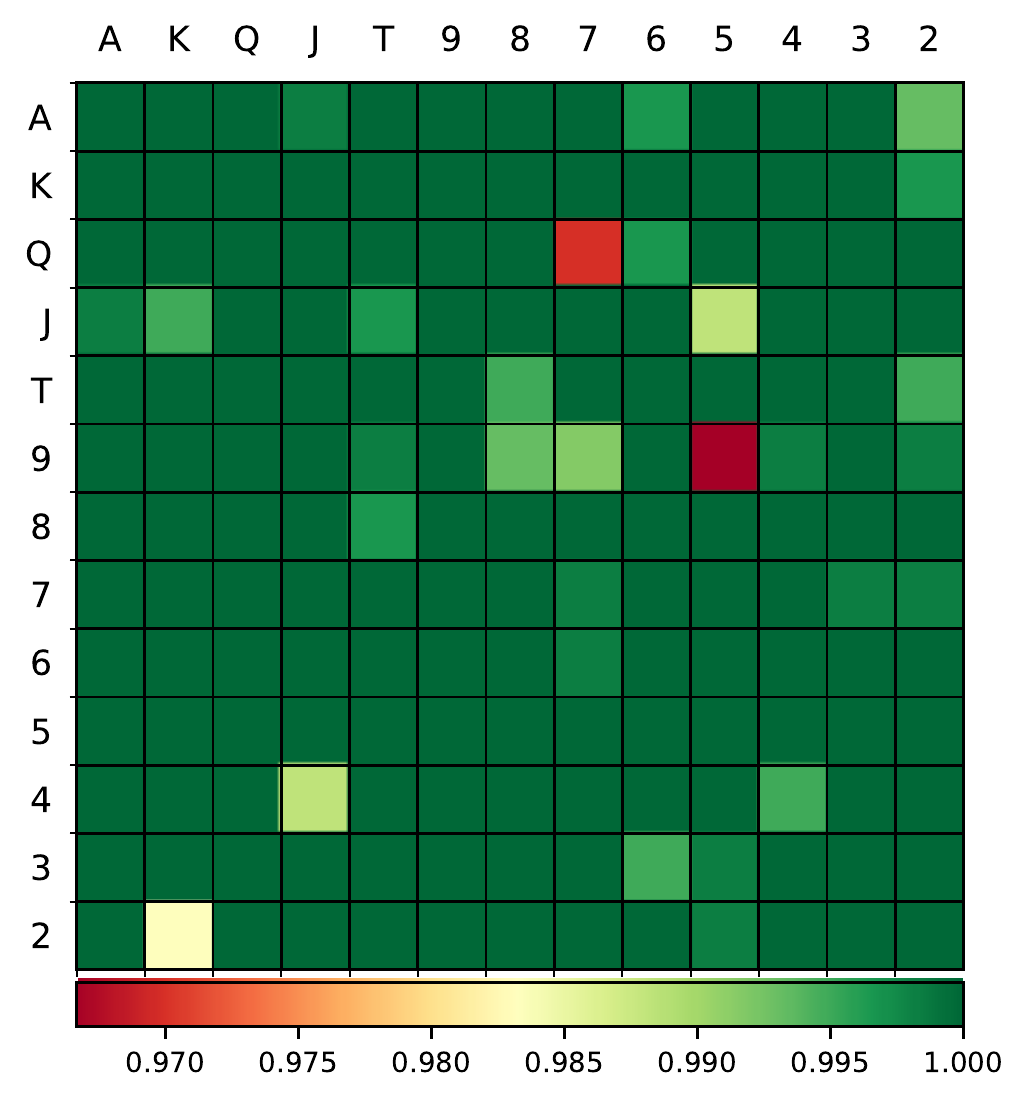}
 \caption{ \label{Flop_evolve_dealer_strategy_B1_P16} 
 Dealer's strategy in section \ref{sec:flop_poker_evolution}, evolved for $(a,B)~=~(8,1)$.
 }
  \end{minipage}
\end{figure}

\begin{figure}[!tbp]
  \centering
  \begin{minipage}[b]{0.4\textwidth}
     \includegraphics[width=7cm,height=7cm,keepaspectratio]{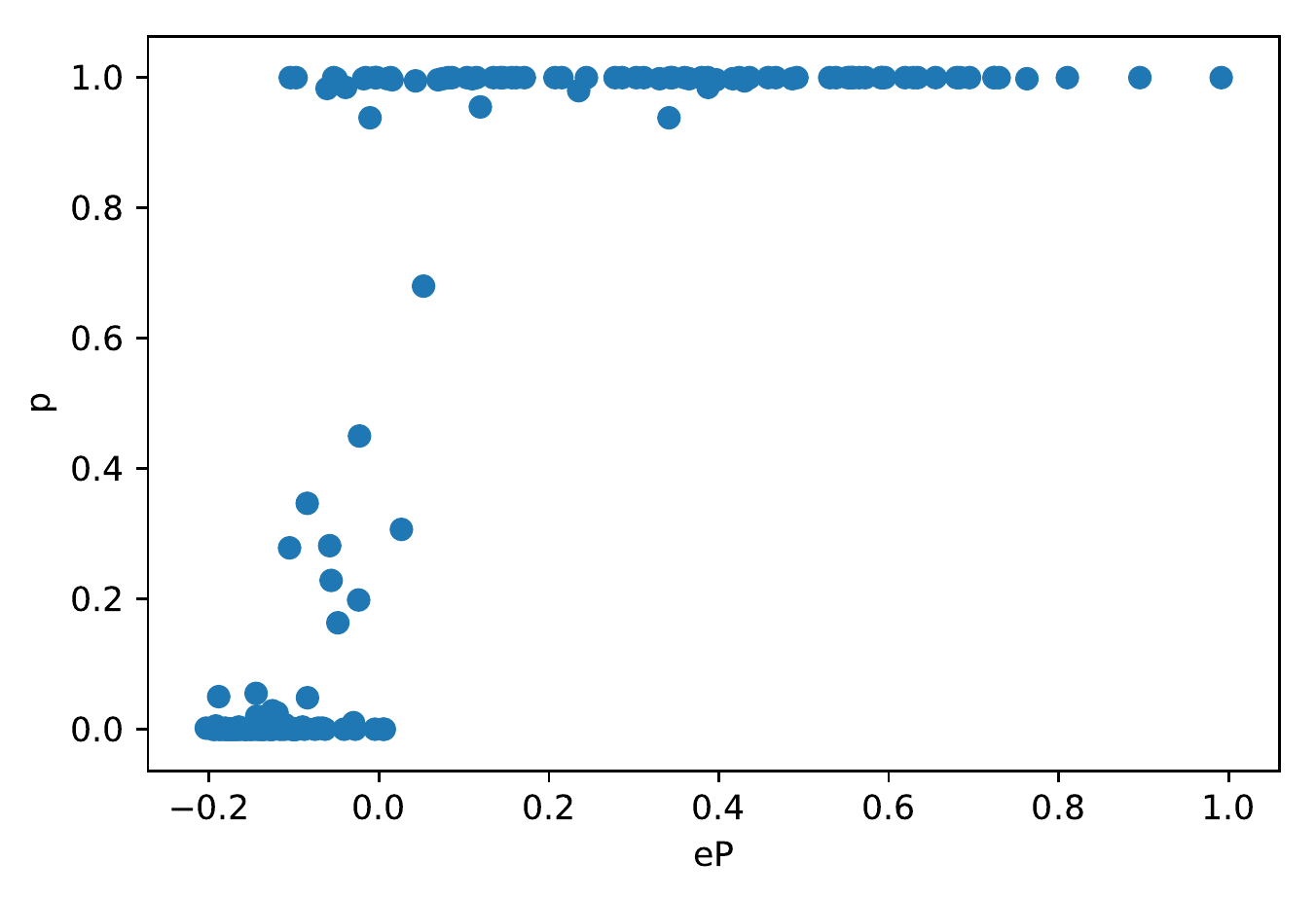}
 \caption{ \label{eP_p_B2_P2} 
 Player's strategy in section \ref{sec:flop_poker_evolution}, evolved for $(a,B)=(1,2)$, vs the corresponding
 value of $e_P(i)$, as defined in (\ref{eP_def}).
 }
  \end{minipage}
  \hfill
  \begin{minipage}[b]{0.4\textwidth}
   \includegraphics[width=7cm,height=7cm,keepaspectratio]{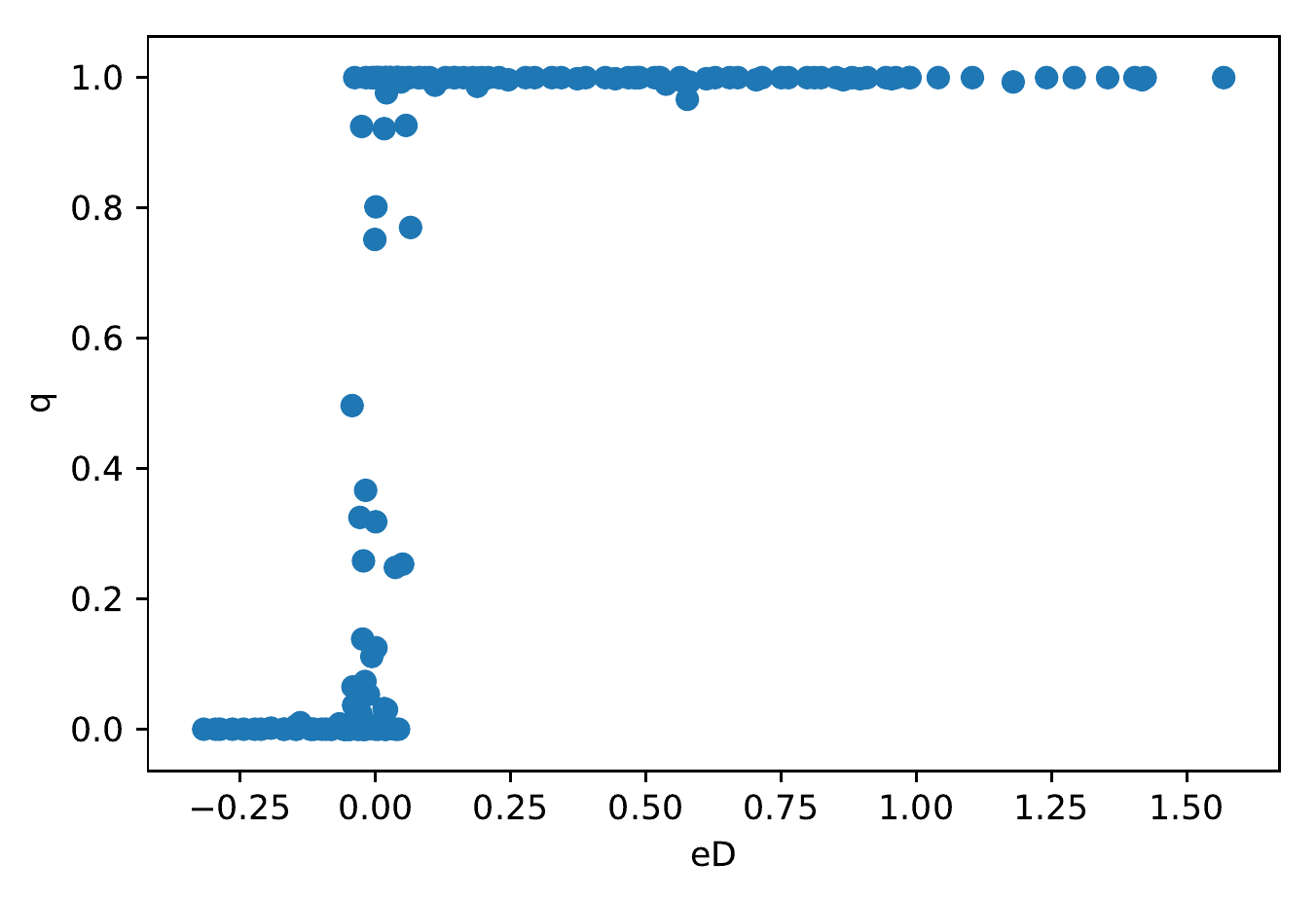}
 \caption{ \label{eD_q_B2_P2} 
 Dealer's strategy in section \ref{sec:flop_poker_evolution}, evolved for $(a,B)=(1,2)$, vs the corresponding
 value of $e_D(j)$, as defined in (\ref{eD_def}).
 }
  \end{minipage}
\end{figure}

\begin{figure}[!tbp]
  \centering
  \begin{minipage}[b]{0.4\textwidth}
     \includegraphics[width=7cm,height=7cm,keepaspectratio]{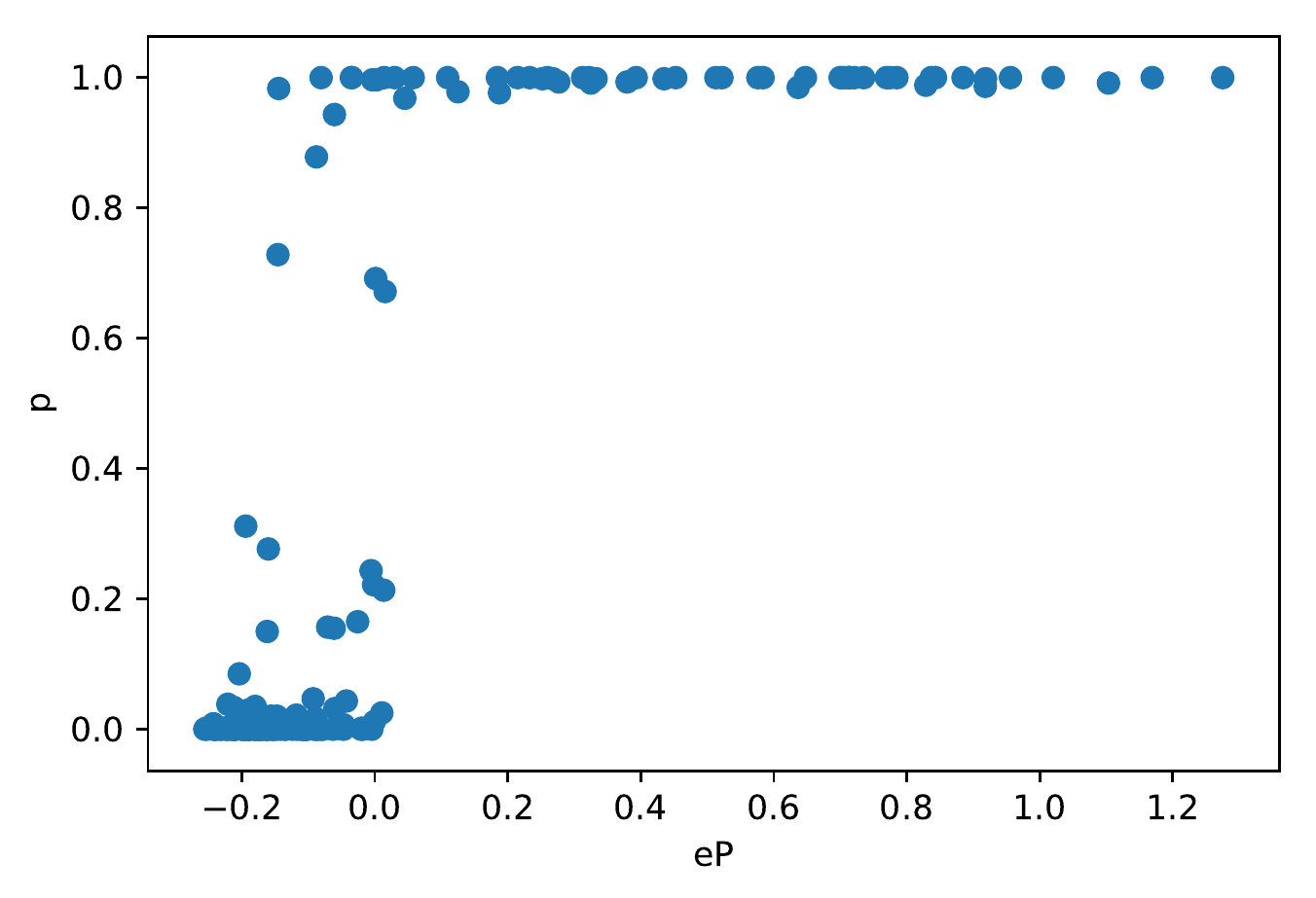}
 \caption{ \label{eP_p_B4_P2} 
 Player's strategy in section \ref{sec:flop_poker_evolution}, evolved for $(a,B)=(1,4)$, vs the corresponding
 value of $e_P(i)$, as defined in (\ref{eP_def}).
 }
  \end{minipage}
  \hfill
  \begin{minipage}[b]{0.4\textwidth}
   \includegraphics[width=7cm,height=7cm,keepaspectratio]{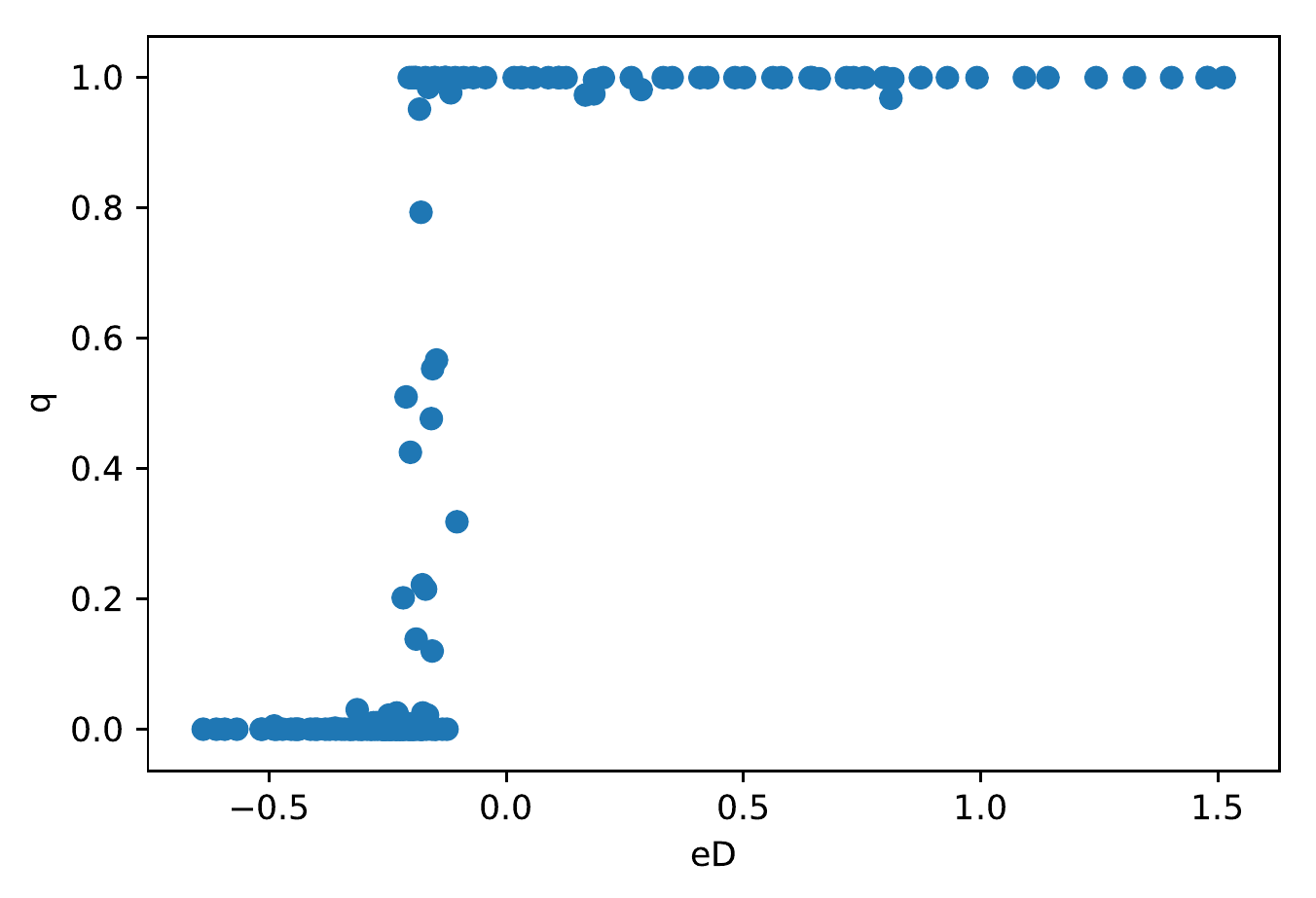}
 \caption{ \label{eD_q_B4_P2} 
 Dealer's strategy in section \ref{sec:flop_poker_evolution}, evolved for $(a,B)=(1,4)$, vs the corresponding
 value of $e_D(j)$, as defined in (\ref{eD_def}).
 }
  \end{minipage}
\end{figure}

\begin{figure}[!tbp]
  \centering
  \begin{minipage}[b]{0.4\textwidth}
     \includegraphics[width=7cm,height=7cm,keepaspectratio]{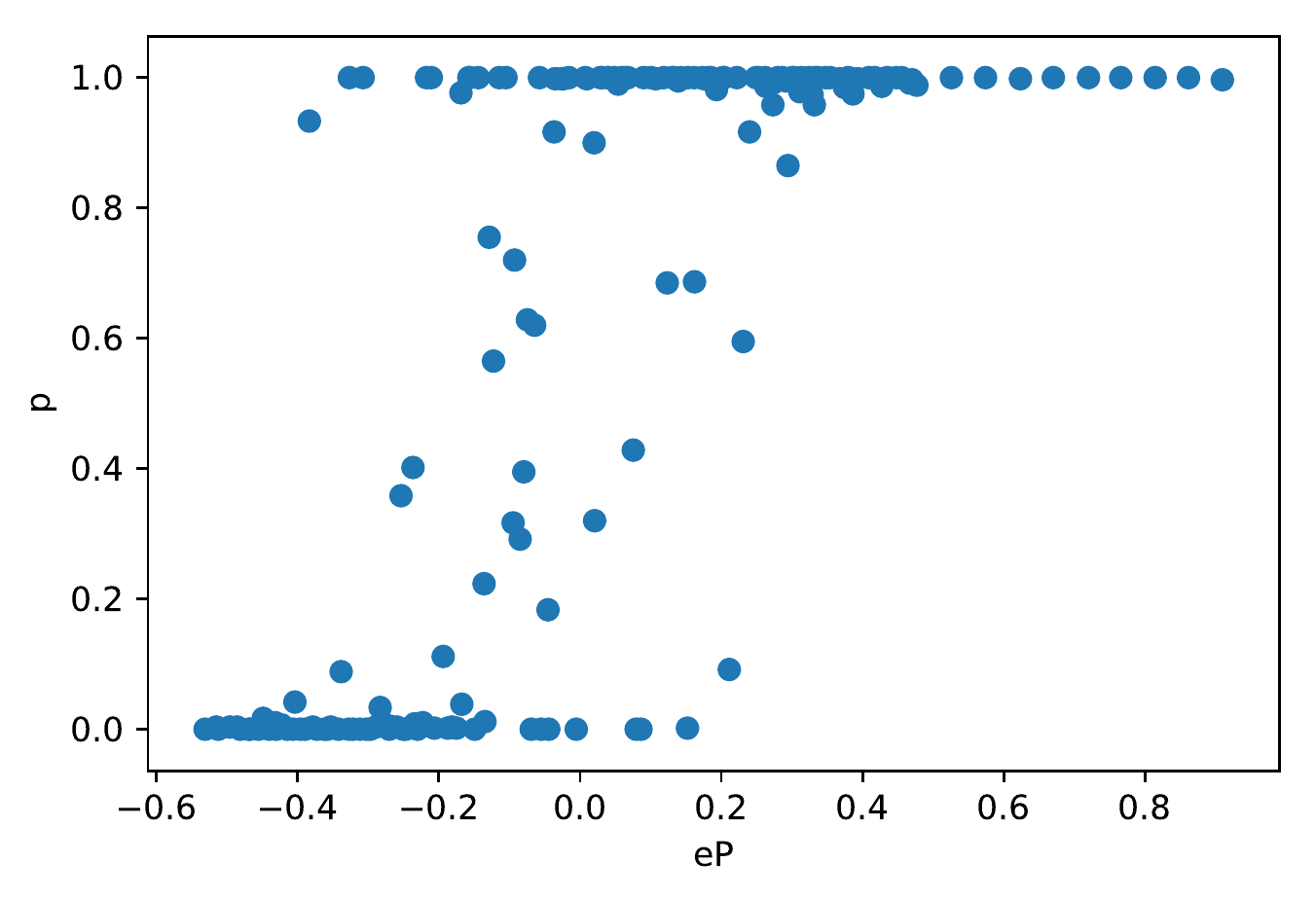}
 \caption{ \label{eP_p_B1_P16} 
 Player's strategy in section \ref{sec:flop_poker_evolution}, evolved for $(a,B)=(8,1)$, vs the corresponding
 value of $e_P(i)$, as defined in (\ref{eP_def}).
 }
  \end{minipage}
  \hfill
  \begin{minipage}[b]{0.4\textwidth}
   \includegraphics[width=7cm,height=7cm,keepaspectratio]{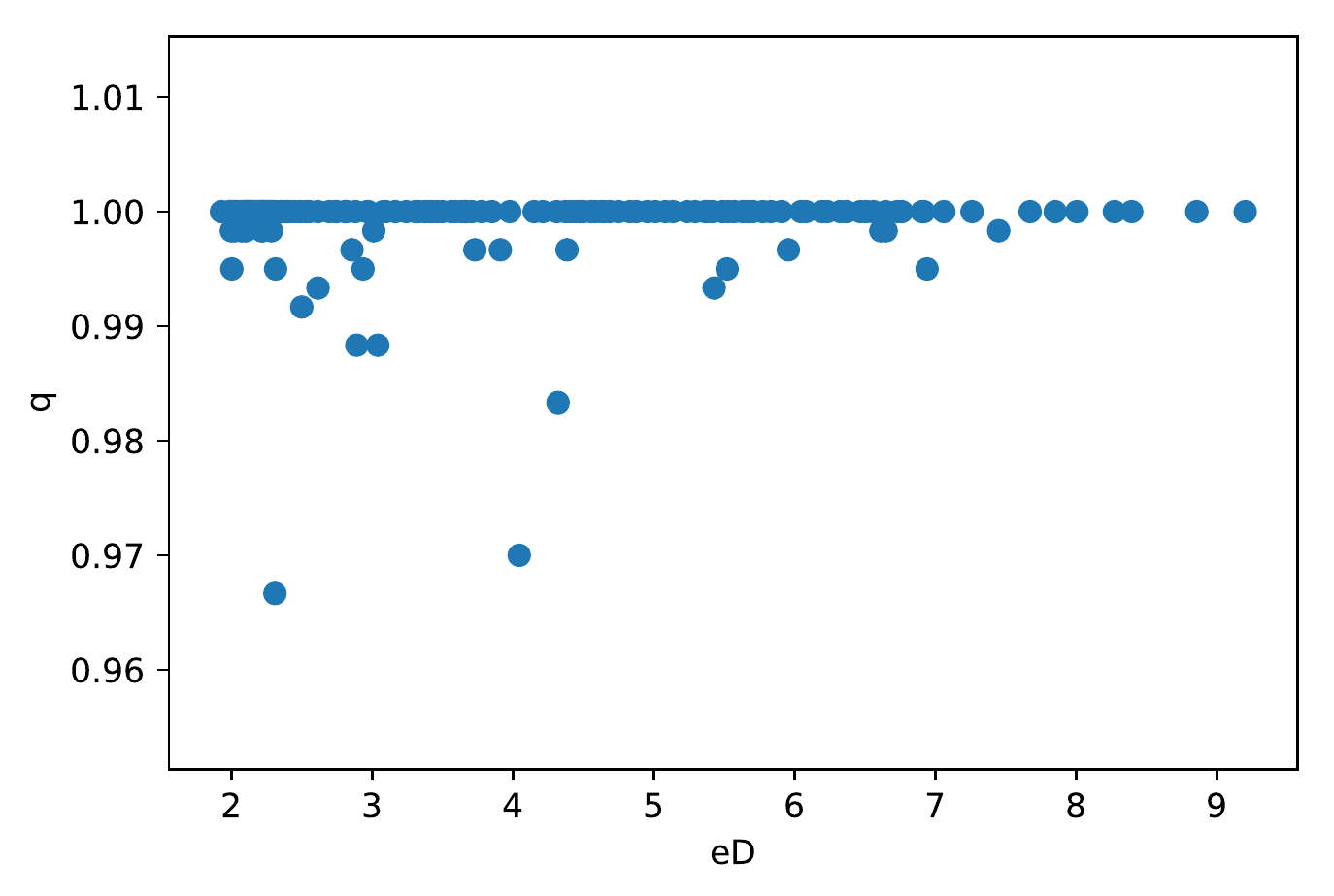}
 \caption{ \label{eD_q_B1_P16} 
 Dealer's strategy in section \ref{sec:flop_poker_evolution}, evolved for $(a,B)=(8,1)$, vs the corresponding
 value of $e_D(j)$, as defined in (\ref{eD_def}).
 }
  \end{minipage}
\end{figure}

In section \ref{sec:flop_poker} we described the rules of the flop poker game,
and derived expressions (\ref{eP_def}), (\ref{eD_def})
which determine the Nash equilibrium strategies $p(i)$, $q(j)$
for the Player and the Dealer. In this section we will use the genetic
 algorithm to derive the approximate Nash equilibrium strategies for the
Player and the Dealer. We will then test the agreement of the genetic algorithm
output with the expressions (\ref{eP_def}), (\ref{eD_def}).

Applying evolutionary algorithms to the game of poker is subtle.
Poker is a non-transitive game, as can be illustrated by the following well-known example
(see \cite{Chen2006} for a recent discussion). Consider the game
in which two players are in a version of the heads up Texas Hold'Em.
Each player can pick one of the three possible two-card private hands: 22, AKo, or JTs.
After one of the players (sucker) makes a pick, the other player (shark) picks one of the two
remaining pairs. Then five community cards are dealt from the remaining deck,
and the player who makes the highest hand wins. These are the probabilities
to win with each of these hands against each of the other hands:
\begin{align}
p\left(22| {\rm AK}_{{\rm o}}\right)&=0.53\,,\qquad  p\left({\rm AK}_{{\rm o}}|22\right)=0.47\,\\
p\left({\rm AK}_{{\rm o}}|{\rm JT}_{{\rm s}}\right)&=0.6\,,
\qquad  p\left({\rm JT}_{{\rm s}}|{\rm AK}_{{\rm o}}\right)=0.39\,\\
p\left({\rm JT}_{{\rm s}}| 22\right)&=0.53\,,\qquad  p\left(22|{\rm JT}_{{\rm s}}\right)=0.46\,.
\end{align}
Therefore regardless of what hand the first player picks, the second player will always
be able to pick a hand which is better on average. 

This is analogous to the the Rock-Paper-Scissors (RPS) game, which is also non-transitive, and therefore
cannot be solved evolutionary (see \cite{MaynardSmith1982} for a discussion
of evolutionary game theory, and the RPS game). 
When applying evolutionary approach to the flop poker game we are hoping that non-transitive
effects, if manifested, give only small fluctuations around the Nash equilibrium.

The flop poker has the same same betting structure as the von Neumann poker.
In the flop poker the strategies for the Player (Dealer) should prescribe
the probabilities to bet (call, if facing bet) for each of $M=169$
possible private hands. The game tree of the flop poker also contains chance nodes indicating possible dealings of the
three community cards. Apart from these distinctions the games of the flop poker and the von Neumann
poker are similar enough, so that we can apply the genetic algorithm
described in subsection \ref{review_evolutionary_programming} to search for the
equilibrium strategy in the flop poker. In particular, each gene in the players strategy
chromosome will take value of either 0 or 1, and the mixed strategy, if it is an equilibrium solution,
is expected to manifest as a population polymorphism.

To improve the evolutionary stability of the equilibrium strategy we will
unite the Player and the Dealer into one Participant player agent, and calculate the
fit of each Participant as the negative squared loss \cite{Quek2009}.
Each Participant therefore carries two chromosomes of the size $M=169$ each,
one encodes the strategy for when the Participant is assigned the first place
in the game (as a Player), and the other encodes the strategy for the second
place in the game (as a Dealer).
The roles of a Player and a Dealer will be assigned
randomly for each Participant.
Then the best
possible fit of each Participant is equal to zero, consistent with the game being zero-sum.
Since the score of the most fit Participants is equal to zero, during the reproduction the parents
are selected with uniform probability. (This is contrasted with the probability proportional
to the positive fitness ofs parents when evolving the strategy in the von Neumann
poker, see section \ref{sec:von_neumann_evolve}.)

In our simulation we start by initializing randomly a population of $N=2000$ Participants.
We evolve the population for $T=1000$ rounds. At each round of evolution $R=10^4$
rounds of the flop poker game take place. Before each round of game all of the $N=2000$
Participants are uniformly shuffled and paired into $N/2=1000$ games.

At the end of each evolution round, after $R=10^4$ rounds of the flop poker game have been
played, the Participants are ranked by their minimized squared loss. Then $\alpha=0.3$
of the most fit participants are selected, and the rest of the Participants are discarded.
The selected Participants replenish the population back to the size $N=2000$
via the two-Participant random breeding with the uniform probability. The mutation probability 
defined in subsection \ref{review_evolutionary_programming} is set to $\pi=10^{-4}$.

We present the results of the evolutionary optimization for $(a,B)=(1,2)$ in figures
\ref{Flop_evolve_player_strategy_B2_P2},
\ref{Flop_evolve_dealer_strategy_B2_P2},
for  $(a,B)=(1,4)$ in figures \ref{Flop_evolve_player_strategy_B4_P2}, \ref{Flop_evolve_dealer_strategy_B4_P2},
and for $(a,B)=(8,1)$
in figures \ref{Flop_evolve_player_strategy_B1_P16}, \ref{Flop_evolve_dealer_strategy_B1_P16}.
We notice that while the genetic algorithm finds the generally correct strategy,
some noise is still present. We can quantify the errors of the evolutionary optimization in the 
following way. We know that the correct equilibrium values of the Player and Dealer
strategies, $p(i)$ and $q(j)$, are determined by the signs of $e_P(i)$, $e_D(j)$,
where the latter are defined by expressions (\ref{eP_def}), (\ref{eD_def}).
Indeed, $e_P(i)>0$ exerts an evolutionary pressure to adapt $p(i)>1$,
while $e_P(i)<0$ favors the adaptation of $p(i)=0$, and similarly for the Dealer's chromosome.
However if the absolute
value of $e_P(i)$ ($e_D(j)$) is small, then the evolutionary pressure on the corresponding $p(i)$
($q(j)$) will be reduced. This effect can be observed by plotting the values of $p(i)$
against $e_P(i)$, and $q(j)$ against $e_D(j)$, see figures \ref{eP_p_B2_P2}, \ref{eD_q_B2_P2} for $(a,B)=(1,2)$,
figures \ref{eP_p_B4_P2}, \ref{eD_q_B4_P2} for $(a,B)=(1,4)$, and
figures \ref{eP_p_B1_P16}, \ref{eD_q_B1_P16} for $(a,B)=(8,1)$.

\section{Counterfactual regret minimization in the flop poker}
\label{sec:flop_poker_cfr}

\begin{figure}[!tbp]
  \centering
  \begin{minipage}[b]{0.4\textwidth}
     \includegraphics[width=7cm,height=7cm,keepaspectratio]{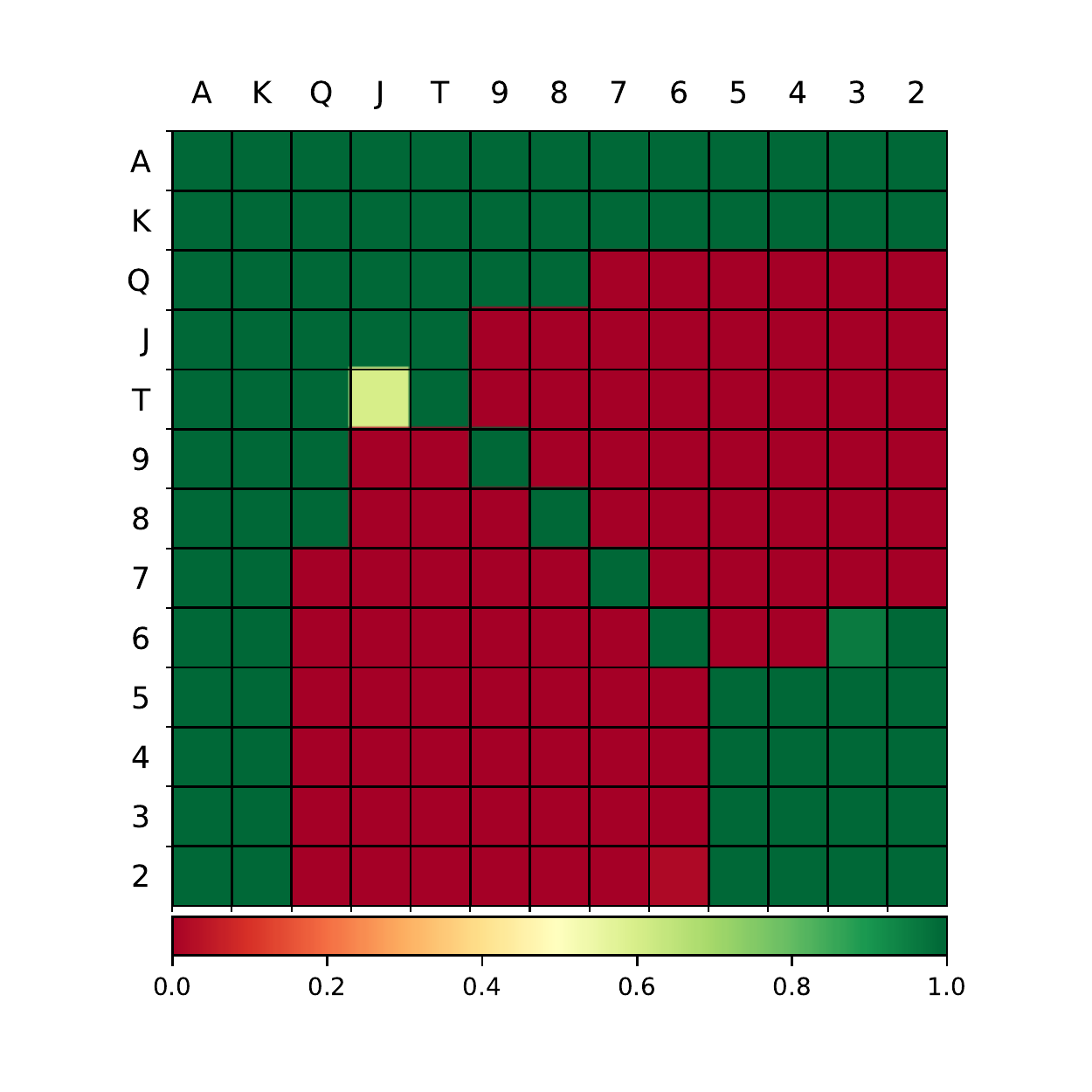}
 \caption{ \label{Flop_regret_player_strategy_B2_P2} 
 Player's strategy in section \ref{sec:flop_poker_cfr} for $(a,B)~=~(1,2)$.
 }
  \end{minipage}
  \hfill
  \begin{minipage}[b]{0.4\textwidth}
   \includegraphics[width=7cm,height=7cm,keepaspectratio]{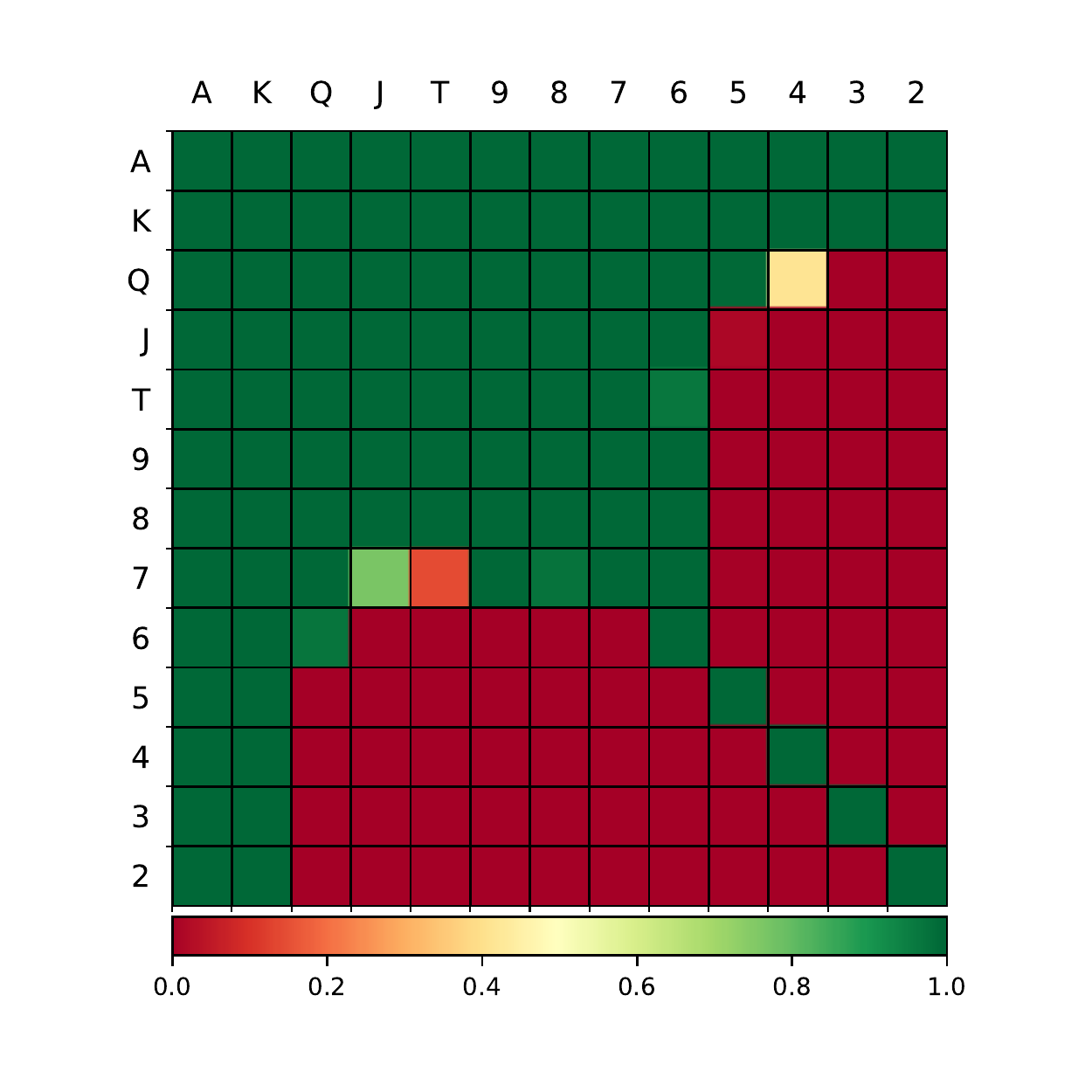}
 \caption{ \label{Flop_regret_dealer_strategy_B2_P2} 
 Dealer's strategy in section \ref{sec:flop_poker_cfr} for $(a,B)~=~(1,2)$.
 }
  \end{minipage}
\end{figure}

\begin{figure}[!tbp]
  \centering
  \begin{minipage}[b]{0.4\textwidth}
     \includegraphics[width=7cm,height=7cm,keepaspectratio]{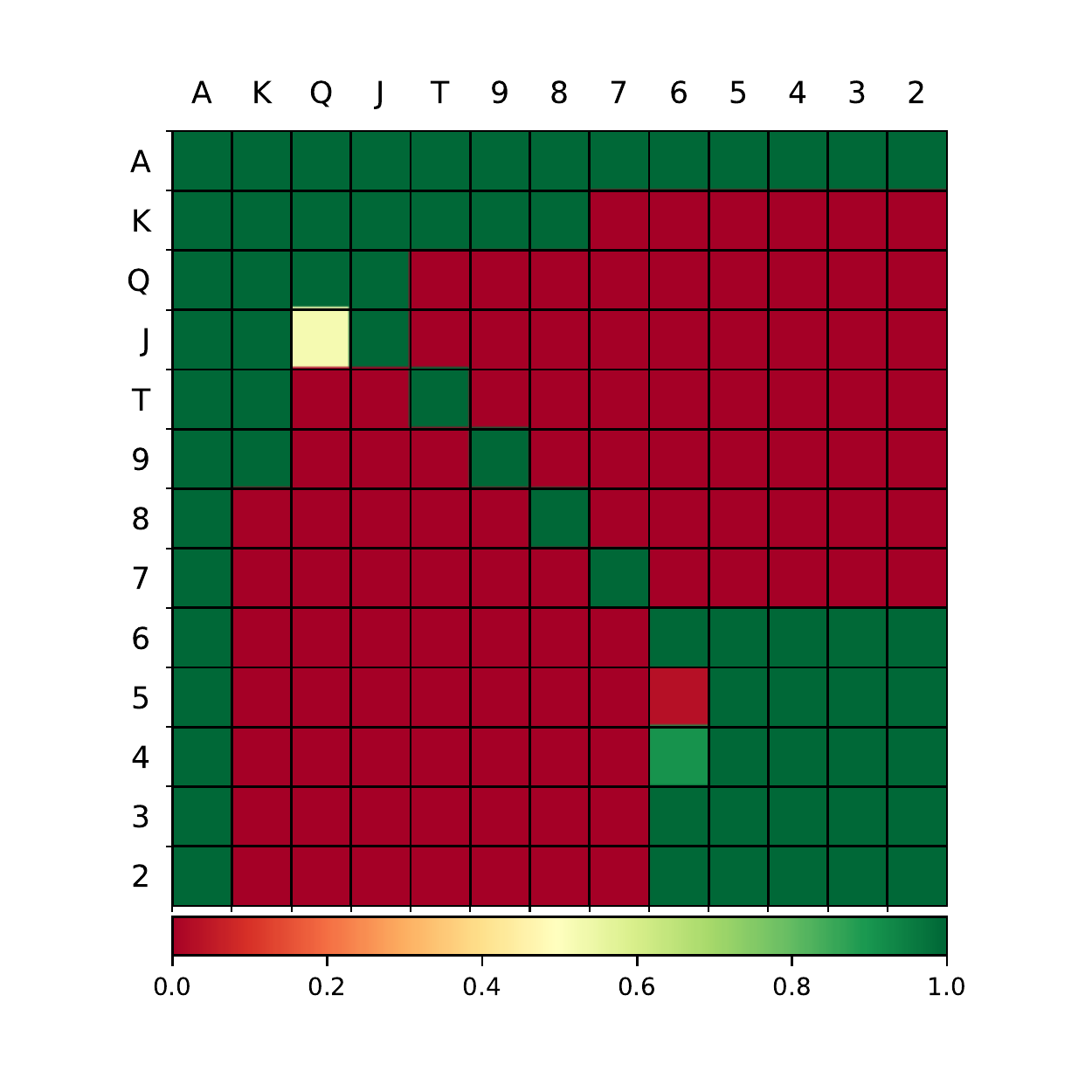}
 \caption{ \label{Flop_regret_player_strategy_B4_P2} 
 Player's strategy in section \ref{sec:flop_poker_cfr} for $(a,B)~=~(1,4)$.
 }
  \end{minipage}
  \hfill
  \begin{minipage}[b]{0.4\textwidth}
   \includegraphics[width=7cm,height=7cm,keepaspectratio]{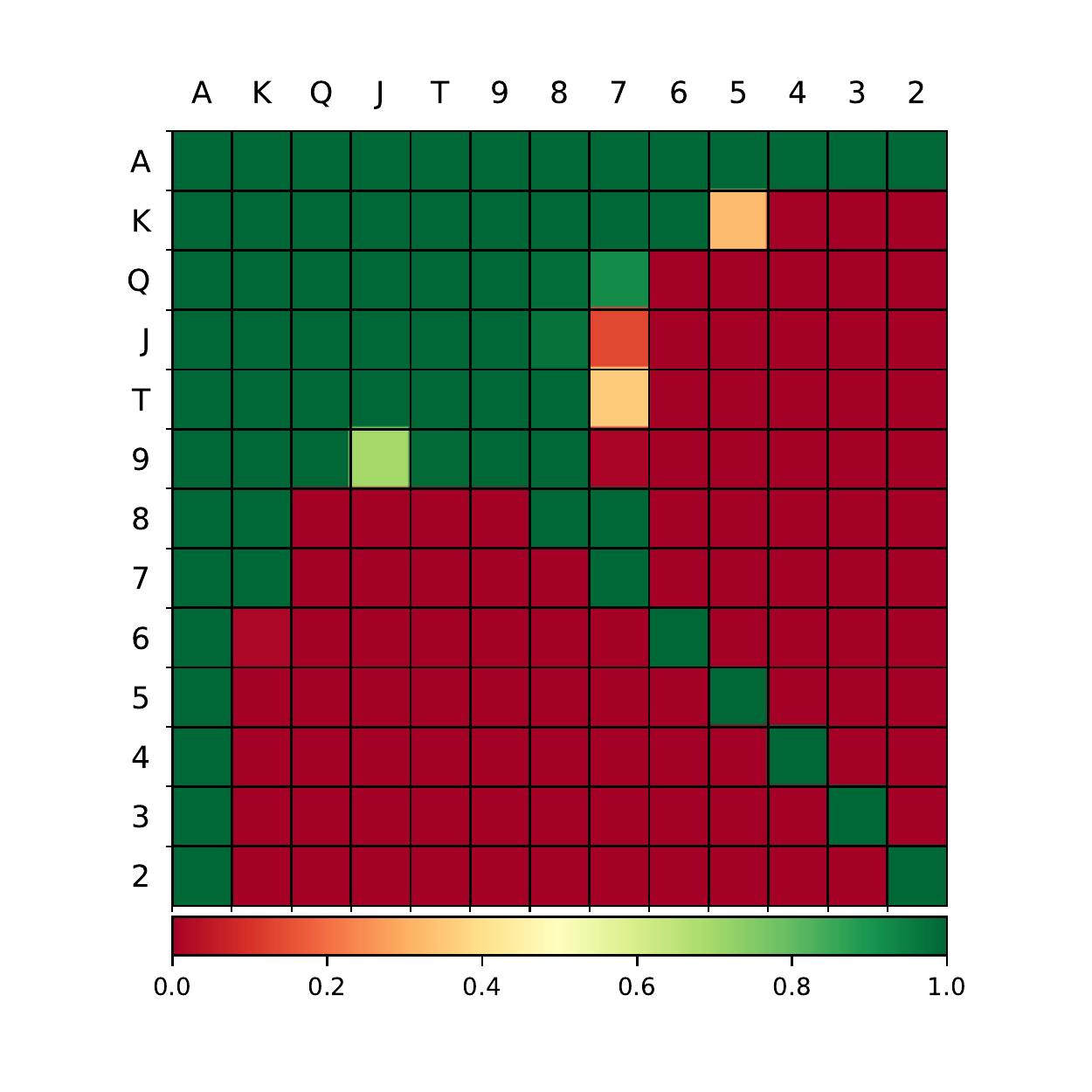}
 \caption{ \label{Flop_regret_dealer_strategy_B4_P2} 
 Dealer's strategy in section \ref{sec:flop_poker_cfr} for $(a,B)~=~(1,4)$.
 }
  \end{minipage}
\end{figure}

\begin{figure}[!tbp]
  \centering
  \begin{minipage}[b]{0.4\textwidth}
     \includegraphics[width=7cm,height=7cm,keepaspectratio]{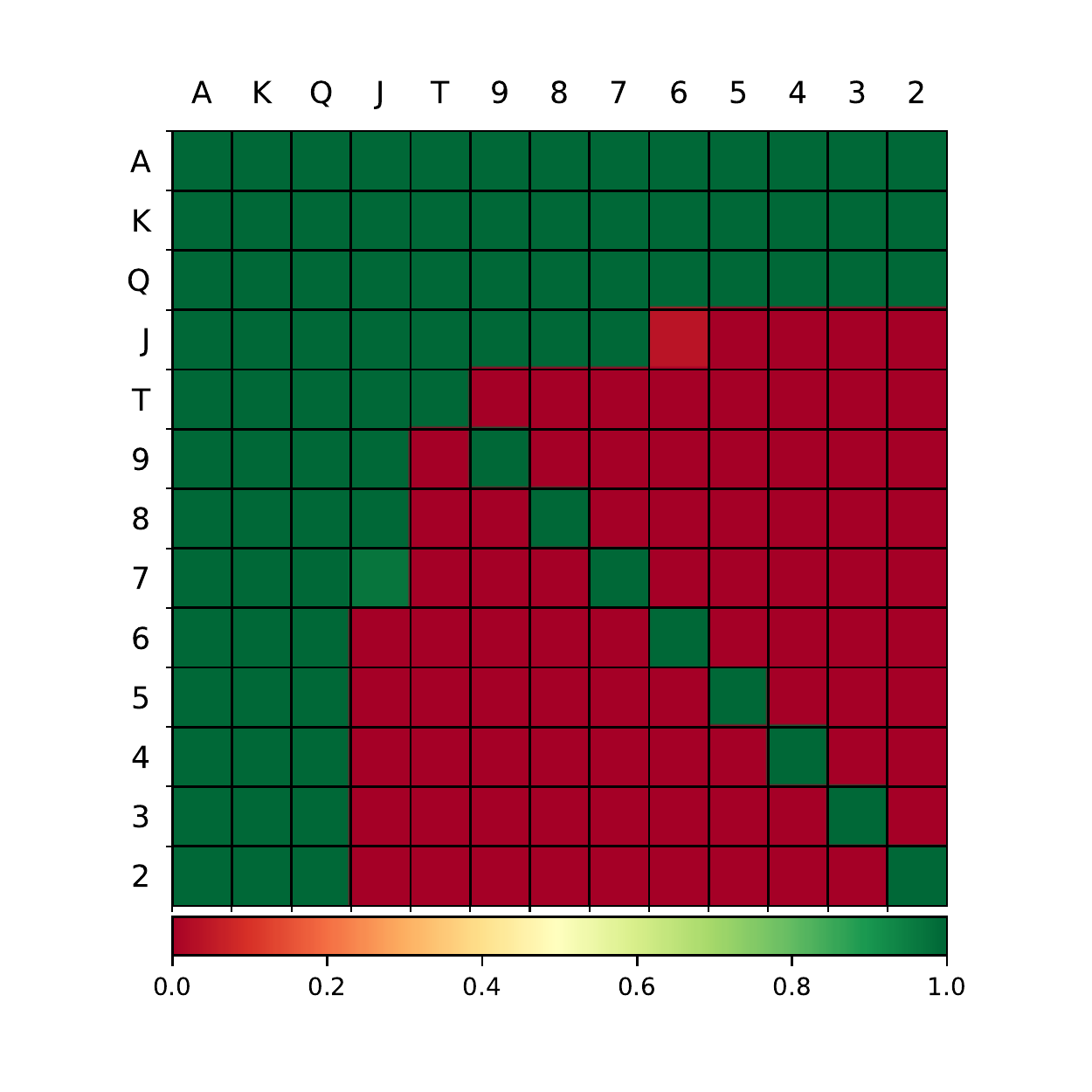}
 \caption{ \label{Flop_regret_player_strategy_B1_P16} 
 Player's strategy in section \ref{sec:flop_poker_cfr} for $(a,B)~=~(8,1)$.
 }
  \end{minipage}
  \hfill
  \begin{minipage}[b]{0.4\textwidth}
   \includegraphics[width=7cm,height=7cm,keepaspectratio]{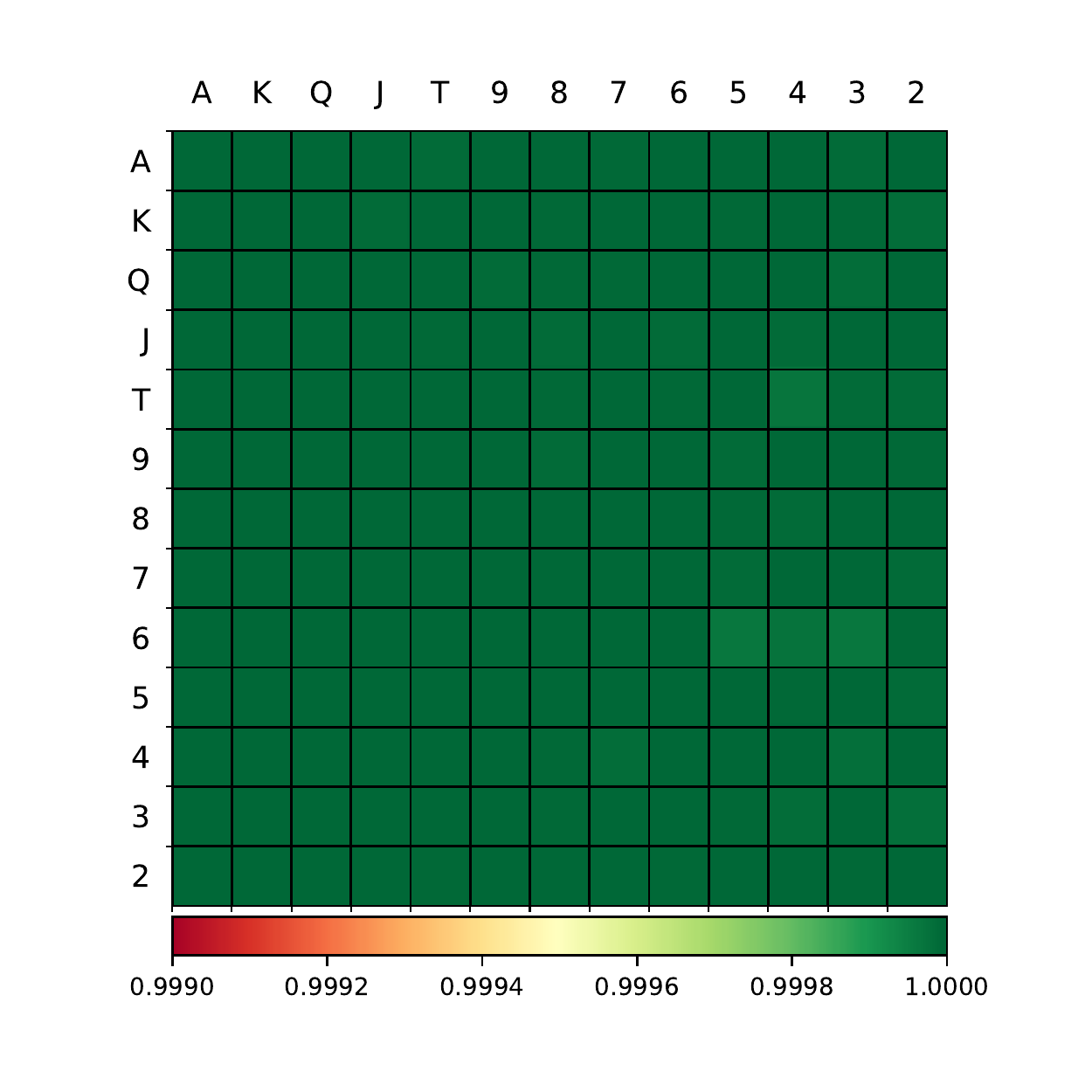}
 \caption{ \label{Flop_regret_dealer_strategy_B1_P16} 
 Dealer's strategy in section \ref{sec:flop_poker_cfr} for $(a,B)~=~(8,1)$.
 }
  \end{minipage}
\end{figure}

\begin{figure}[!tbp]
  \centering
  \begin{minipage}[b]{0.4\textwidth}
     \includegraphics[width=7cm,height=7cm,keepaspectratio]{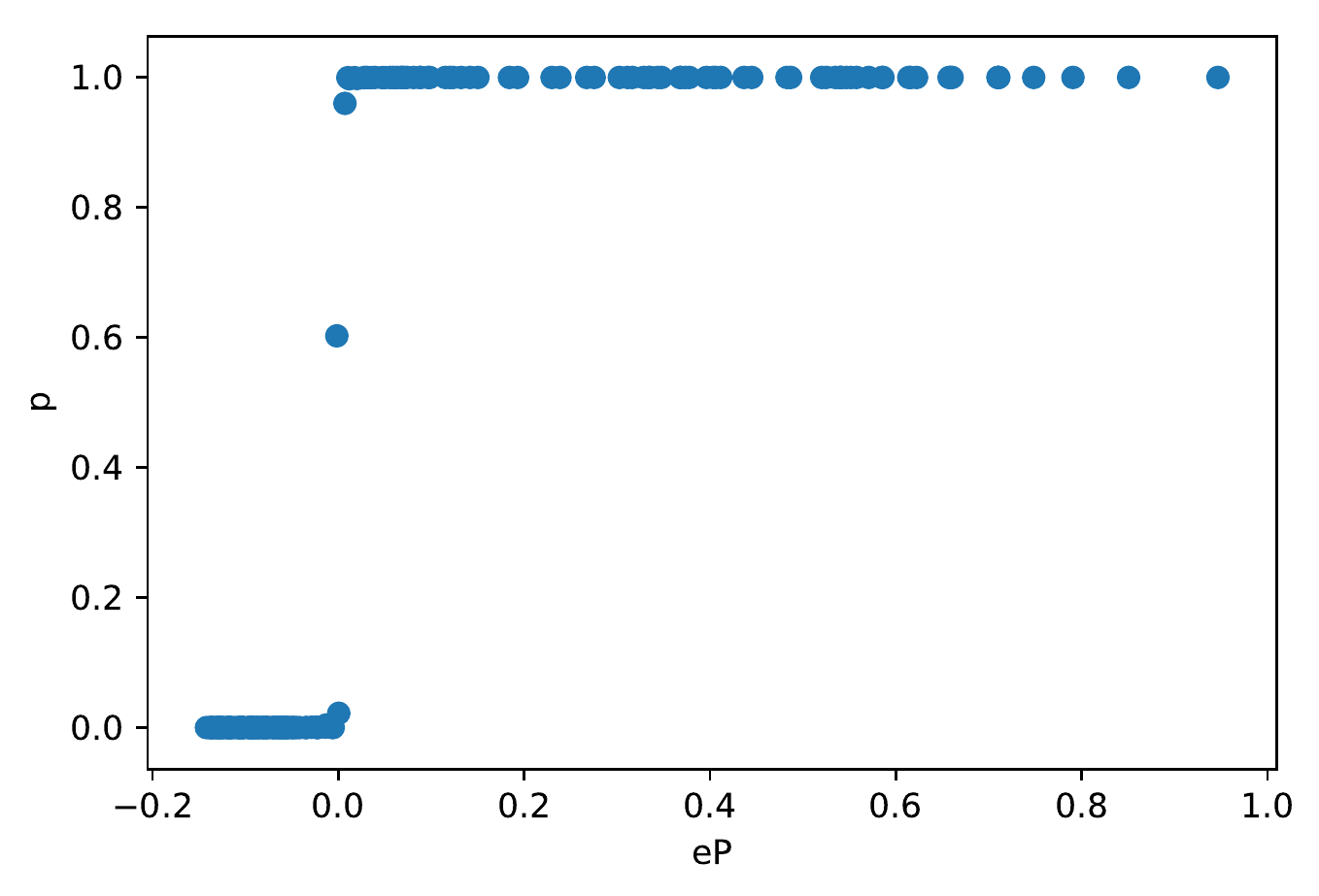}
 \caption{ \label{eP_p_B2_P2_regret} 
 Player's strategy in section \ref{sec:flop_poker_cfr} for $(a,B)=(1,2)$, vs the corresponding
 value of $e_P(i)$, as defined in (\ref{eP_def}).
 }
  \end{minipage}
  \hfill
  \begin{minipage}[b]{0.4\textwidth}
   \includegraphics[width=7cm,height=7cm,keepaspectratio]{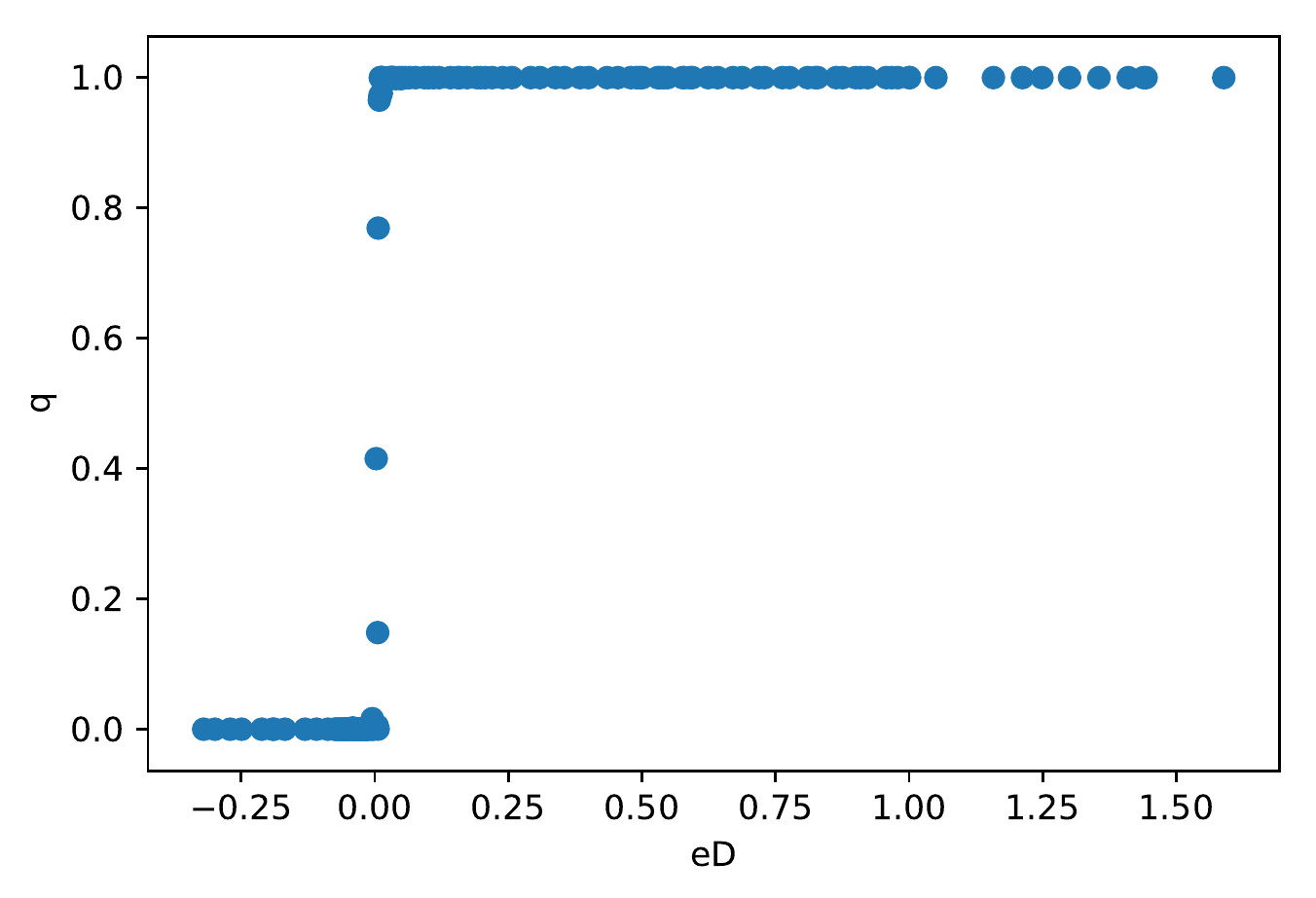}
 \caption{ \label{eD_q_B2_P2_regret} 
 Dealer's strategy in section \ref{sec:flop_poker_cfr} for $(a,B)=(1,2)$, vs the corresponding
 value of $e_D(j)$, as defined in (\ref{eD_def}).
 }
  \end{minipage}
\end{figure}

\begin{figure}[!tbp]
  \centering
  \begin{minipage}[b]{0.4\textwidth}
     \includegraphics[width=7cm,height=7cm,keepaspectratio]{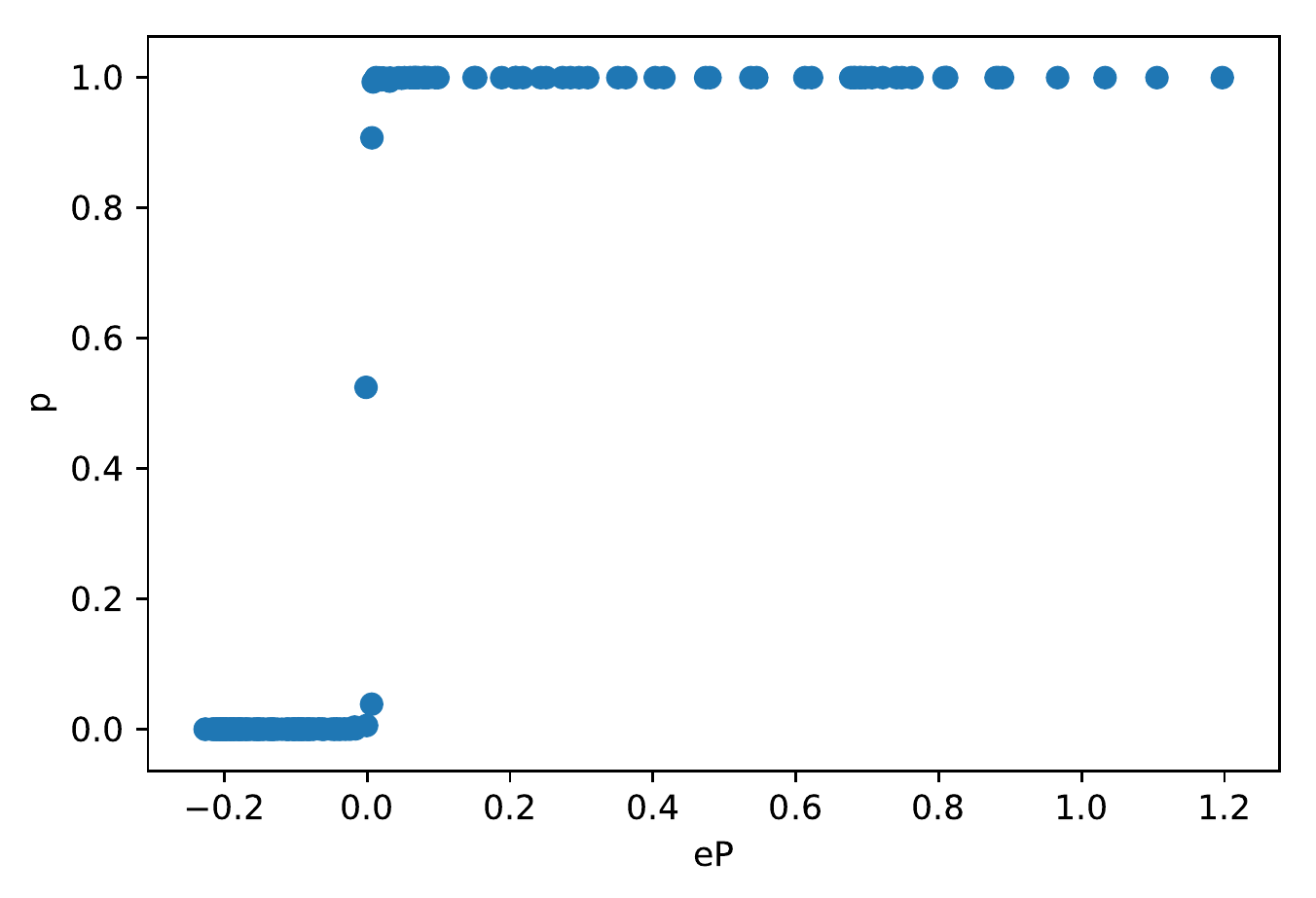}
 \caption{ \label{eP_p_B4_P2_regret} 
 Player's strategy in section \ref{sec:flop_poker_cfr} for $(a,B)=(1,4)$, vs the corresponding
 value of $e_P(i)$, as defined in (\ref{eP_def}).
 }
  \end{minipage}
  \hfill
  \begin{minipage}[b]{0.4\textwidth}
   \includegraphics[width=7cm,height=7cm,keepaspectratio]{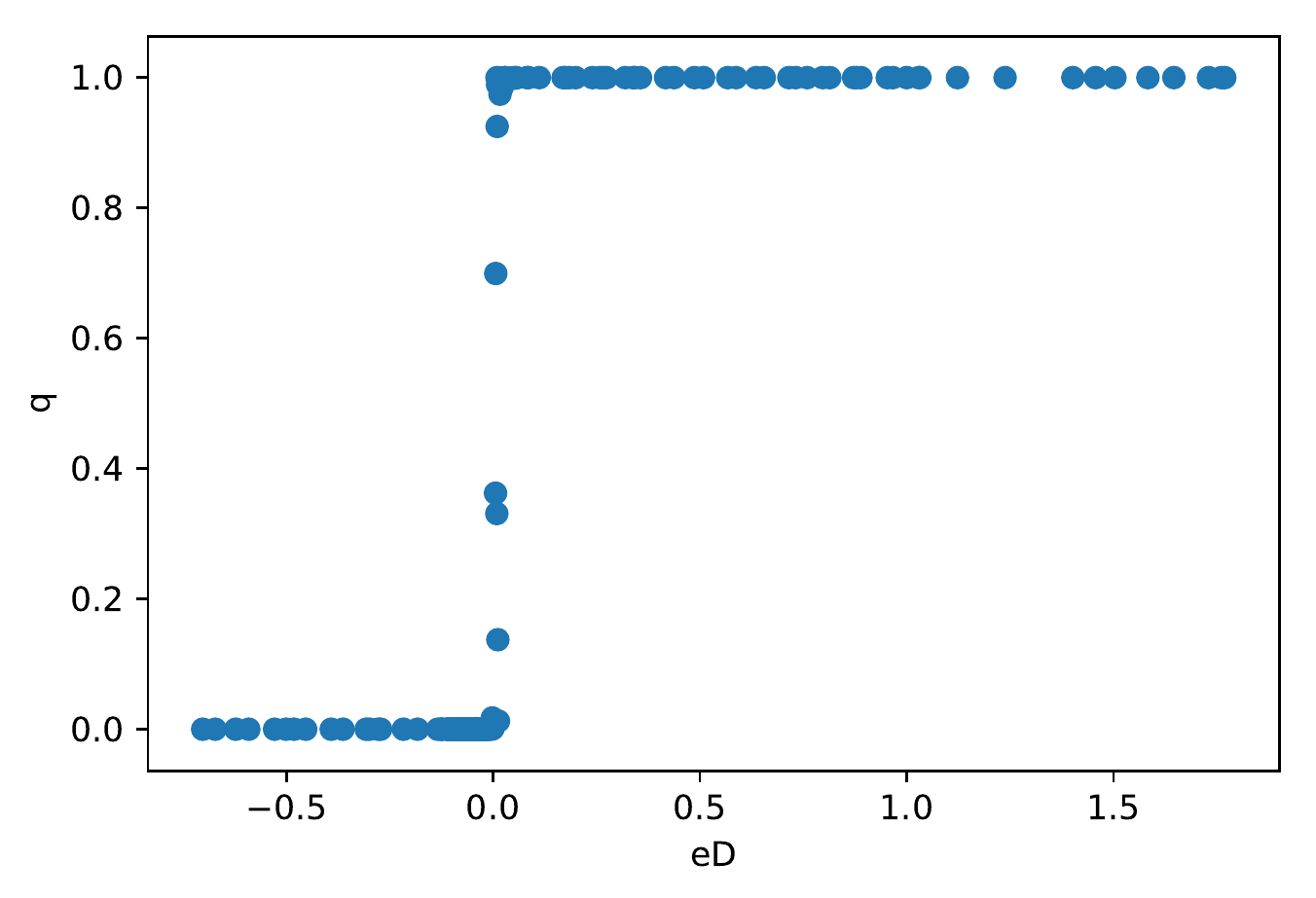}
 \caption{ \label{eD_q_B4_P2_regret} 
 Dealer's strategy in section \ref{sec:flop_poker_cfr} for $(a,B)=(1,4)$, vs the corresponding
 value of $e_D(j)$, as defined in (\ref{eD_def}).
 }
  \end{minipage}
\end{figure}

\begin{figure}[!tbp]
  \centering
  \begin{minipage}[b]{0.4\textwidth}
     \includegraphics[width=7cm,height=7cm,keepaspectratio]{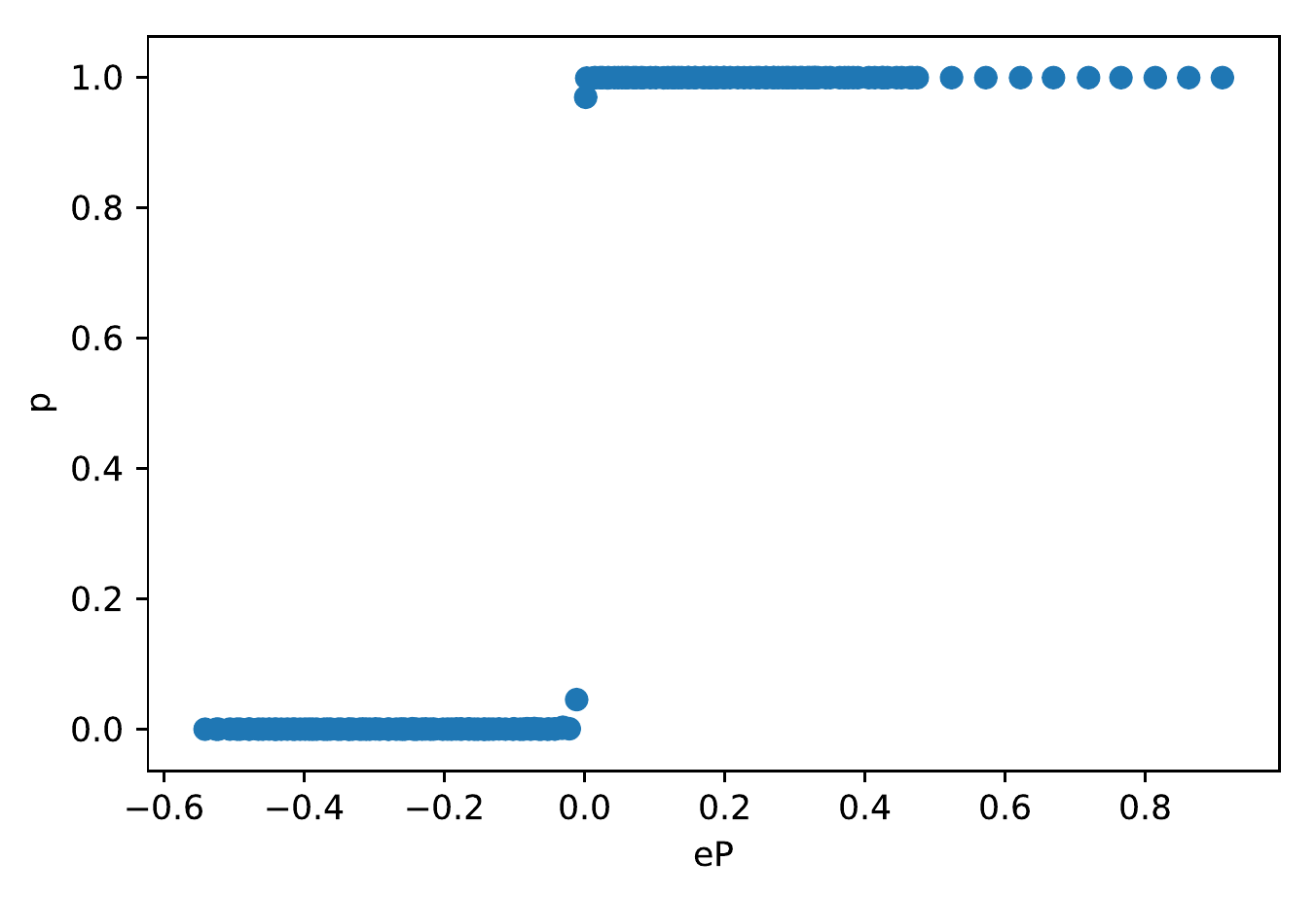}
 \caption{ \label{eP_p_B1_P16_regret} 
 Player's strategy in section \ref{sec:flop_poker_cfr} for $(a,B)=(8,1)$, vs the corresponding
 value of $e_P(i)$, as defined in (\ref{eP_def}).
 }
  \end{minipage}
  \hfill
  \begin{minipage}[b]{0.4\textwidth}
   \includegraphics[width=7cm,height=7cm,keepaspectratio]{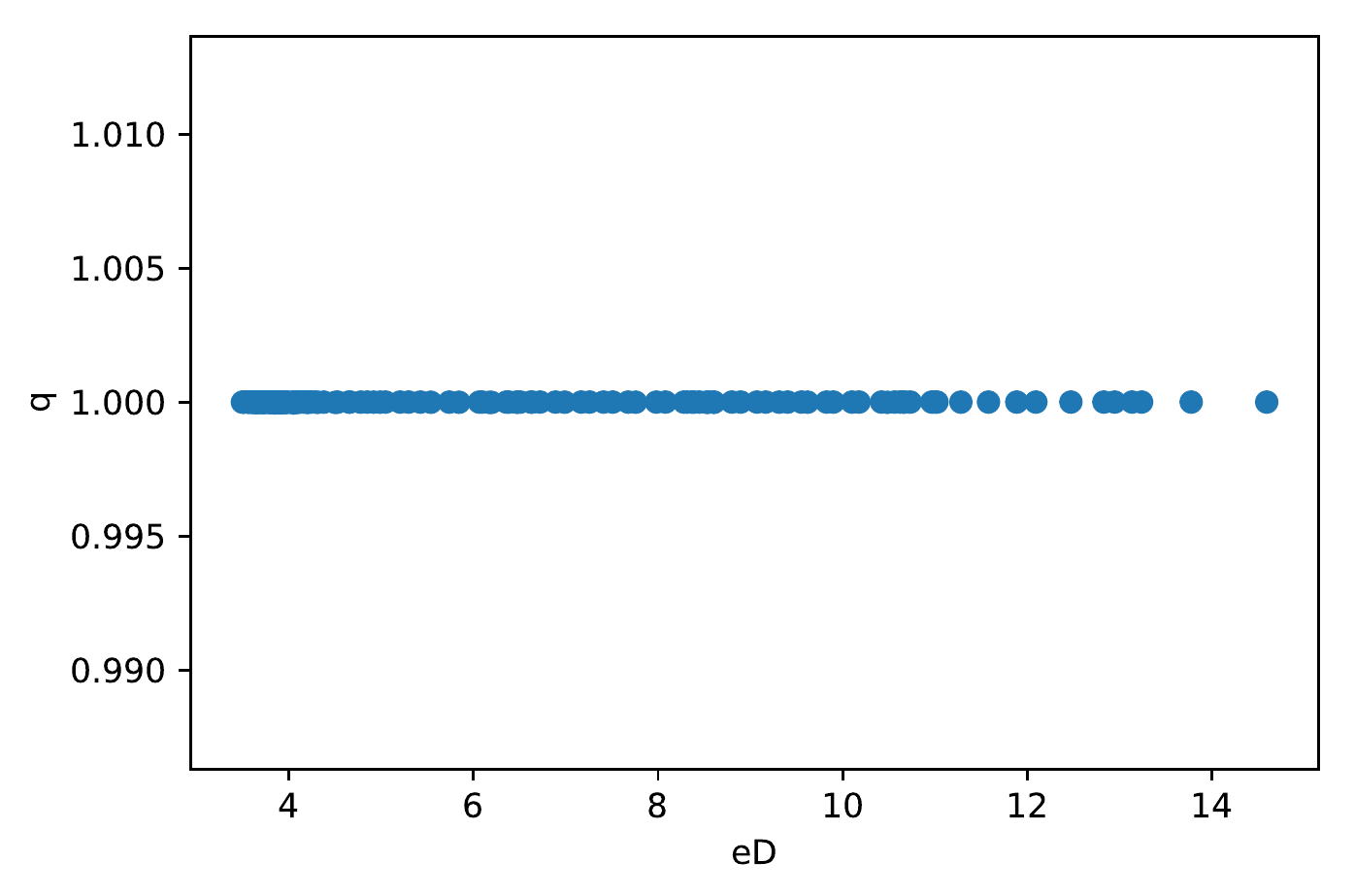}
 \caption{ \label{eD_q_B1_P16_regret} 
 Dealer's strategy in section \ref{sec:flop_poker_cfr} for $(a,B)=(8,1)$, vs the corresponding
 value of $e_D(j)$, as defined in (\ref{eD_def}).
 }
  \end{minipage}
\end{figure}

In subsection \ref{CFR_review} we reviewed the counterfactual regret minimization algorithm
on the example of the von Neumann poker. 
The decision nodes of the flop poker game tree have a similar structure to the game tree of the
von Neumann poker, as discussed in section \ref{sec:flop_poker_evolution}. Therefore we can adapt the CFR
algorithm described in subsection \ref{CFR_review} to search for the Nash equilibrium strategies
in the flop poker. We provide the resulting strategy after $T=2\times 10^9$ game rounds for $(a,B)=(1,2)$
in figures \ref{Flop_regret_player_strategy_B2_P2}, \ref{Flop_regret_dealer_strategy_B2_P2},
and for $(a,B)=(1,4)$ in figures \ref{Flop_regret_player_strategy_B4_P2},
 \ref{Flop_regret_dealer_strategy_B4_P2}. We also provide the resulting strategy
after $T=10^8$ game rounds for $(a,B)=(8,1)$ in figures \ref{Flop_regret_player_strategy_B1_P16},
\ref{Flop_regret_dealer_strategy_B1_P16}.

We notice that these results are similar to the results of the evolutionary optimization
given in section \ref{sec:flop_poker_evolution}. Since for (almost) every $i,j=1,\dots,169$ the equilibrium
Player/Dealer strategy in the flop poker is pure, we notice that the CFR algorithm gives
a more precise result than the genetic  algorithm.
To illustrate performance of the CFR algorithm
we plot the values of $p(i)$
against $e_P(i)$ (defined in (\ref{eP_def})), and $q(j)$ against $e_D(j)$ (defined in (\ref{eD_def})),
see figures \ref{eP_p_B2_P2_regret},
\ref{eD_q_B2_P2_regret} for $(a,B)=(1,2)$,
figures \ref{eP_p_B4_P2_regret}, \ref{eD_q_B4_P2_regret} for $(a,B)=(1,4)$, and
figures \ref{eP_p_B1_P16_regret}, \ref{eD_q_B1_P16_regret} for $(a,B)=(8,1)$.

We also calculate the game value for the CFR strategies,
\begin{align}
E_P(a=1,b=2)&=0.15\,,\qquad E_D(a=1,b=2)=-0.15\,,\\
E_P(a=1,b=4)&=0.14\,,\qquad E_D(a=1,b=4)=-0.14\,,\\
E_P(a=8,b=1)&=0.11\,,\qquad E_D(a=8,b=1)=-0.11\,.
\end{align}
Notice that when the pot gets large it becomes optimal for the Dealer to always
call, and for the Player to never bluff. This is because for the small bet to ante ratio
the odds are good enough for the Dealer to be risking a loss of extra bet.
If the Dealer always calls then it is no longer advantageous for the Player to bluff.

\section{Discussion}
\label{sec:discussion}

In this paper we compared performance of the genetic algorithm
and the counterfactual regret minimization algorithm in the task of calculating the near-equilibrium
strategies in the von Neumann poker, and the simplified Texas Hold'Em poker (referred to as the flop poker).
Von Neumann poker features a continuum of possible equilibrium strategies for the player in the second
position in the game. Only one of these strategies is admissible, that is, maximally exploitative
(among all the other equilibrium strategies) of the opponent's deviations from the Nash equilibrium.
Importantly,
we demonstrated that the genetic algorithm finds the equilibrium strategy in the von Neumann poker
which is also (approximately) admissible, while the counterfactual regret minimization algorithm finds one of the possible
equilibrium strategies at random.

We also demonstrated that the genetic algorithm has weaker convergence properties than the counterfactual
regret minimization algorithm, when applied to the game of the flop poker. We suggested that this
behavior of the genetic algorithm can be explained by the reduced selection pressure acting
on the particular strategy entries (corresponding to the genes of the strategy chromosomes, in the 
context of genetic algorithms). On the other hand the counterfactual regret minimization algorithm
predicts the correct Nash equilibrium strategies with a high degree of accuracy.  
It would be interesting to further investigate how performance
of the genetic algorithm can be improved in order to decrease the noise in the evolved strategy output.
However, as discussed, the use of genetic algorithms for the poker games should
be taken with caution due to the non-transitive nature of the game.

The results of this paper have been obtained using the code written by the author in C++ and Python.
The original code is partly available on the author's GitHub page, and can also be requested from the author.

\section*{Appendix. Details of the poker hand evaluator}
\label{sec:appendix}

To obtain results in sections \ref{sec:flop_poker_evolution} and \ref{sec:flop_poker_cfr} we used
numerical methods of genetic algorithms and counterfactual regret minimization,
involving ranking of a large number of five-card poker hands. In this appendix we describe the
efficient poker hand evaluator which we used.

As discussed in subsection \ref{introducing_von_neumann} all five-card poker hands can be ranked
in strength, with the strength rank taking one of 7462 values.
Therefore to determine the winning player we need to calculate the  ranks for each player, and see
which player has the winning rank. In notations of subsection \ref{introducing_von_neumann} the winning rank
has the smallest value, with rank 1 standing for the Royal Flush, and rank 7462 standing for
the High Card 7~5~4~3~2.

The most straightforward way to proceed with construction of the hand evaluator is therefore
to use a deck of card objects, where each card object has attributes of rank and suit. Then
for each poker hand of five cards we first determine whether it is a flush or not. If the hand
is a flush, its rank will be searched for in the table of flushes, which includes the Straight Flush and the non-straight
Flush hands. If the hand is not a flush, its rank will be searched for in the table of non-flushes, which contains the 
Four of a Kind, Full House, non-flush Straight, Three of a Kind, Two and One Pair, and High Card.
The reason for separating flushes and non-flushes into different tables is because from the point of view
of the unique (that is, non-degenerate w.r.t. equivalent in strength choices of suits)
ranks composition of the High Card hands are the same as the non-straight Flush 
hands, and the Straight Flush hands are the same as the non-flush Straight hands. 

This method can be optimized even further, by using a deck of cards where each card, instead of being
an object with the attributes of rank and suit, is represented by a unique integer number. Inspired by
\cite{CactusKev} we represent each card as a number $c=s+r$, where $s$ takes one of the 
values 256, 512, 1024, 2048, representing the suits,
and $r$ takes one of the values  2, 3, 5, 7, 11, 13, 17, 19, 23, 29, 31, 37, 41 representing the ranks.
The numbers $s$ encoding suits are chosen so that their binary representation
(assignment of the suit names is arbitrary, the bold-faced bits are reserved for the rank $r$)
\begin{align}
s_{{\rm spades}}=256&=0{\rm b} \, 0001{\bf 00000000}\\
s_{{\rm hearts}}=512&=0{\rm b} \, 0010{\bf 00000000}\\
s_{{\rm clubs}}=1024&=0{\rm b} \, 0100{\bf 00000000}\\
s_{{\rm diamonds}}=2048&=0{\rm b} \, 1000{\bf 00000000}
\end{align}
will make it easy to check whether the hand is a flush or not. This is because we can perform
a fast bitwise shift operation $\gg$ on the (copies of) cards $c_i$ in the hand $\{c_i\}_{i=1}^{i=5}$,
killing off the first eight bits (to get the ranks $r_i=c_i-s_i$
out of the way), and then take a bitwise product of all the resulting numbers in the hand.
The only time the result will be a non-zero number is when the hand is a flush. 
This will tell
us whether we should be searching for the hand rank in the table of flushes or the table of non-flushes.
Those tables in turn encode unique products of prime numbers $\prod _{i=1}^5r_i$
for all possible hands in there, where $r_i$
can be extracted by taking a bitwise product of each $c_i$ with 255.

\section*{Acknowledgment}

I would like to thank M.~Baggioli, B.~Galilo and A.~Teimouri for numerous discussions, and A.~Teimouri for
collaboration on parts of this project, and reading and commenting on the draft of this paper.

\end{document}